\documentclass[format=acmsmall, review=false, screen=true]{acmart}

\usepackage{amsmath}

\usepackage{booktabs} 
\usepackage{algorithmic}
\usepackage[ruled]{algorithm2e} 

\usepackage[caption=false,font=footnotesize]{subfig}
\usepackage{balance}
\usepackage{amssymb}
\usepackage{todonotes}
\usepackage[english]{babel}
\usepackage{breqn}
\usepackage{textcomp} 

\SetAlFnt{\small}
\SetAlCapFnt{\small}
\SetAlCapNameFnt{\small}
\SetAlCapHSkip{0pt}
\IncMargin{-\parindent}

\setcopyright{acmlicensed}

\acmDOI{0000001.0000001}

\received{February 2007}
\received[revised]{March 2009}
\received[accepted]{June 2009}

\begin{document}
\title[DNA Molecular Storage System through Bacterial Nanonetworks]{DNA Molecular Storage System: Transferring Digitally Encoded Information through Bacterial Nanonetworks}

\author{Federico Tavella}
\affiliation{%
  \institution{University of Padua}
  \department{Department of Mathematics}
  \city{Padua}
  \country{Italy}}
\email{federico.tavella@studenti.unipd.it}

\author{Alberto Giaretta}
\affiliation{%
  \institution{\"{O}rebro University}
  \department{Department of Science and Technology, Centre for Applied Autonomous Sensor Systems}
  \city{\"{O}rebro}
  \country{Sweden}}
\email{alberto.giaretta@oru.se}

\author{Triona Marie Dooley-Cullinane}
\affiliation{%
  \institution{Waterford Institute}
  \department{Pharmaceutical \& Molecular Biotechnology Research Centre, Department of Science}
  \city{Waterford}
  \country{Ireland}}
\email{tmdooley-cullinane@wit.ie}

\author{Mauro Conti}
\affiliation{%
  \institution{University of Padua}
  \department{Department of Mathematics}
  \city{Padua}
  \country{Italy}}
\email{conti@math.unipd.it}

\author{Lee Coffey}
\affiliation{%
  \institution{Waterford Institute}
  \department{Pharmaceutical \& Molecular Biotechnology Research Centre, Department of Science}
  \city{Waterford}
  \country{Ireland}}
\email{lcoffey@wit.ie}

\author{Sasitharan Balasubramaniam}
\affiliation{%
  \institution{Tampere University of Technology}
  \department{Department of Electronic and Communication Engineering}
  \city{Tampere}
  \country{Finland}}
\affiliation{%
  \institution{Waterford Institute of Technology}
  \department{Telecommunications Software \& Systems Group (TSSG)}
  \city{Waterford}
  \country{Ireland}}
\email{sasi.bala@tut.fi}


\begin{abstract}
Since the birth of computer and networks, fuelled by pervasive computing and ubiquitous connectivity, the amount of data stored and transmitted has exponentially grown through the years. Due to this demand, new solutions for storing data are needed, and one promising media is the DNA. This storage solution provides numerous advantages, which includes the ability to store dense information while achieving long-term stability. However, the question as how the data can be retrieved from a DNA-based archive, still remains.

In this paper, we aim to address this question by proposing a new storage solution that relies upon molecular communication, and in particular bacterial nanonetworks. Our solution allows digitally encoded information to be stored into non-motile bacteria, which compose an archival architecture of clusters, and to be later retrieved by engineered motile bacteria, whenever reading operations are needed.

We conducted extensive simulations, in order to determine the reliability of data retrieval from non-motile storage clusters, placed at different locations. Aiming to assess the feasibility of our solution, we have also conducted wet lab experiments that show how bacteria nanonetworks can effectively retrieve a simple message, such as "Hello World", by conjugation with non-motile bacteria, and finally mobilize towards a final point.
\end{abstract}

%
%

\ccsdesc[500]{Computing methodologies~Modeling and simulation}
\ccsdesc[500]{Computer systems organization~Other architectures}
\ccsdesc[500]{Hardware~Communication hardware, interfaces and storage}
\ccsdesc[500]{Hardware~Emerging technologies}

%
%

\keywords{DNA Encoding, Data Storage, Bacterial Nanonetworks, Molecular Communications.}

\maketitle

\renewcommand{\shortauthors}{Tavella et al.}

\section{Introduction}
%
%
%
%
Worldwide, the quantity of new data is rapidly increasing on a daily basis, due to the massive number of connected devices to the Internet. Indeed, many of the current software features that we are provided nowadays, such as navigation systems and social networks, heavily rely upon machine learning techniques which run over big data. These techniques create even more data, which leads to the increased demanding of storage that we experience since the last decade. To face this shortage, new infrastructures like data centres~\cite{Eco1} and cloud computing infrastructures \cite{Eco2} are ceaselessly built; as an example, as part of \emph{Internet.org} initiative, in 2013 Facebook built a new high capacity data center~\cite{Facebook}. 

Besides the large-scale infrastructure, new solutions at the hardware level (HDDs, SSDs, servers and clusters) are also required. 
In recent years, besides silicon technology to store digital information, researchers investigated alternative storage media, and one example is the \emph{Deoxyribonucleic Acid}~(\emph{DNA}). DNA, which is naturally found inside biological cells, encodes information that represents functionalities and characteristics of the cell itself: from a computer science perspective, where the cell represents the hardware, the DNA represents the software. 

Since the '80s, techniques for encoding DNA has continually improved in terms of performance, overcoming even Moore's Law\cite{Microsoft}. A good example of Digital DNA encoding was recently proposed by Goldman et al.~\cite{Goldman}. However, a major question still remains as to how we can possibly automate the \texttt{Reading} process of this encoded information, especially if they are archived collectively. In this paper, we aim to address this \texttt{Reading} process through a bio-inspired communication mechanism, known as \emph{molecular communications}.


In this paper, bacterial nanonetworks with engineered \emph{E. Coli} pick up the plasmids with digitally encoded DNA information from an archive library, in order to deliver the information to a point where sequencing operations can run to decode it. Through simulations, we assess how much the movement randomness affects the engineered bacteria while mobilizing towards a destination point. Our aim is to determine the whole system reliability applied to tasks of information retrieval, by conjugation with clusters of motility-restricted bacteria.

The engineered motility is based on the positioning technique proposed by Moore et al.~\cite{Moore11}~\cite{Moore13}, and we adopt a triangulation process that is inspired by the one commonly used in cellular mobile networks. The triangulation of mobile devices is obtained by measuring radio signal strength, while in this paper the trilateration is based on sensing chemical signals emitted by beacons. Our proposal relies on engineered bacteria ability to sense chemicals with different variations, as proposed in~\cite{Moore13}, in order to mobilize towards various spatial points of the archive. Each point reflects the location of a cluster of motility-restricted bacteria, where the desired information can be picked up by motile bacteria through conjugation. 


\subsection{Contribution}
The contributions of this paper are manifold:
\begin{itemize}
\item {\bf Molecular Positioning System (MPS)}: we propose a positioning system that enables the engineered bacteria to sense chemoemissions and, by programming different types of receptors, to mobilize towards a specific location described by a predefined concentration of chemicals;
\item {\bf Digitally enocded-DNA archive system}: we propose a way to enable bacteria to recover digital encoded DNA information from an archive system, and deliver such information to a target destination for later sequencing;
\item {\bf System validation}: we conduct a set of simulations to assess the precision of MPS, as well as the efficiency of DNA pick-up tasks, in order to evaluate the whole system performance;
\item {\bf Wet lab experiments}: finally, we also run a thorough set of wet lab experiments to demonstrate the feasibility of using bacteria to pick up information, digitally encoded into plasmids, from motile-restricted bacteria, and deliver this information to a destination point. 
\end{itemize}

\subsection{Organization}
This paper is organized as follows: Section~\ref{sec:related} briefly discusses the current state of art about DNA storage systems. Section~\ref{sec:architecture} shows the architecture of our model and describes how the MPS is modelled and implemented, while Section~\ref{sec:our_encoding} how data are encoded into DNA. Section~\ref{sec:simulation} discusses our simulations and evaluates the related results. In Section~\ref{sec:wetexp} we describe how we conducted wet lab experiments to demonstrate the feasibility of our proposal. Finally, Section~\ref{sec:conclusion} presents our conclusions.


\section{Related Work}
\label{sec:related}

The field of molecular communications aims to develop artificial communication system from biological components that are found in nature~\cite{Guney12}~\cite{Akyildiz}~\cite{Nakano}. 
Developing biological artificial communication systems at miniature scale can open numerous opportunities such as advanced healthcare solutions, as well as environmental monitoring and protection. In order to realize a fully functional network, a certain degree of control and engineering at nanoscale is required and this could be achieved through {\bf Synthetic Biology}. The community has also proposed an extension of the Internet of Things that we know today as the {\bf Internet of Bio-NanoThings}~({\bf BNTs})~\cite{Akyildiz2}, where miniaturized engineered biological cells can transmit information to the cyberworld. 

One particular form of molecular communication that has received considerable interest is known as \emph{bacterial nanonetworks}, where engineered bacteria such as \emph{E. Coli} uses its motility property as information carriers. Researchers proposed bacteria-based models for molecular communications due to bacteria inherent properties, such as the ability to communicate and signal between each other, as well as mobilizing within an aqueous environment. For example, Balasubramaniam et al.~\cite{Sasi} proposed a multi-hop communication model, using bacteria as carriers for information encoded into plasmids to communicate between multiple nanomachines. Moore et al. proposed positioning systems for BNTs~\cite{Moore11}~\cite{Moore13}, where the relative position could be inferred by concentration gradients. Other examples of proposed applications include engineering BNTs for cooperative target localization~\cite{Okaie} and related techniques for countering security threats~\cite{Giaretta}. 

Given the popularity of digital encoding of DNA, a number of different works have been proposed both for simulations and wet lab experiments to demonstrate the feasibility of the idea. The University of Washington~\cite{Microsoft} developed an architecture for DNA-Encoding and Archiving system, which is structured as a key-value store that uses random access through the \emph{Polymerase Chain Reaction} (\emph{PCR}) technique. The storage system is composed of a DNA synthesizer responsible for encoding data, storage container divided in compartments that store DNA pools, and a DNA sequencer, which reads DNA sequences and converts them back to binary data. In order to implement key-based retrieving mechanism, researchers use selective DNA amplification with PCR.	Taking advantage of DNA sequencing primers, they are able to amplify only strands corresponding to desired data, discarding the undesired parts. In a similar way, using primers enables this technique to give each strand a key before putting it into a pool. The encoding technique uses both a simple 2 bits-1 base match, which is possible from the four different nucleotides (A, C, G, T). 

However, due to the large amount of errors that can result from the sequencing and synthesis, another proposed solution was the 3 based encoding using Huffman encoding combined with a rotational digit to base conversion~\cite{Microsoft}. Other researchers~\cite{Yazdi} focused on a DNA storage method that allows random-access to information encoding and rewriting. Each data block, of length 1000 \emph{bpm}, contains an addressing block at the beginning and end, while the remaining 960 \emph{bpm} is used for text encoding. The rewriting technique is based on replacing certain known blocks of information from the dictionary, and replacing with new words that are to be updated into the storage. In~\cite{Grass}, the encoding work focuses on the longevity of the storage, in particular to handle the errors and this is achieved through physical storage that can provide maximum stability. The storage medium is based on silica, where predictions have been made that this could last for over 2 millions years. 

Blawat et al.~\cite{Blawat} developed a Forward Error Correction technique that could be used to counter various errors that occur from DNA encoding. These errors may range from the maximum run-length, which is the maximum length of identical nucleotides that is only limited to three same nucleotides, or errors that can emerge from deletion or insertion of nucleotides. 
Another research~\cite{Suyehira} presented an algorithm where information encoding and decoding of digital DNA can be performed into Nucleic Acid Memory (NAM). This process is achieved through protein translation that maps the three nucleotides in a row, also known as codon, to hexadecimal characters. The mapping process ensures that minimum errors result from the arrangement of the nucleotides, ensuring stability of the encoded information. A study~\cite{Shipman} has demonstrated that storing DNA vectors into bacteria can ensure high reliability. Researchers showed a recording system that uses the nucleotide content, temporal ordering, in order to encode arbitrary information within the genomes of a population of cells, and has shown its reliability.

None of the related works examine in depth how encoded data can be transferred between different points and, at the same time, how to ensure reliability by storing such data in a biological cell. We try to address each of these issues through our proposal.

\section{DNA Archival - System Architecture}
\label{sec:architecture}
Apart from encoding data within DNA, the question as how to retrieve such information has to be addressed by any kind of proposed storage system. We envision a system where bacteria are the main actors, both for storage and reading purposes. Indeed, not only data is stored within motile-restricted bacteria, but also motile bacteria are responsible for moving towards pre-defined points, read the desired information through bacterial conjugation and, finally, deliver such information to an endpoint.

\begin{figure}[ht]
\center
\includegraphics[trim=0mm 0mm 0mm 0mm,clip,width=0.8\textwidth]{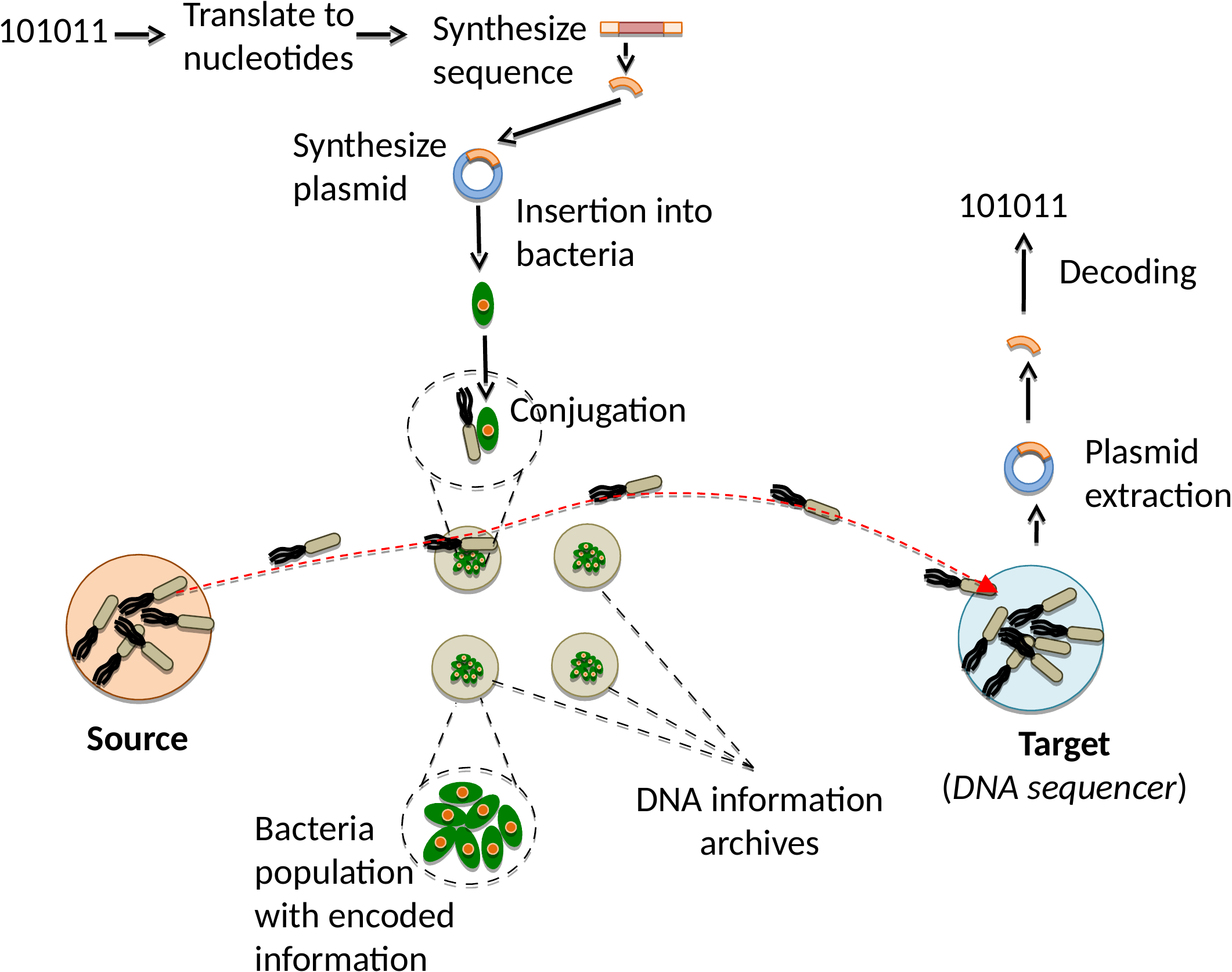}
\caption{Overall system architecture of the DNA archival system that enables reading using bacterial nanonetworks. Motile bacteria (in grey) swim from the Source towards each of the archive points, which contain different information. Reached the motile-restricted bacteria, conjugation starts to retrieve the encoded information contained into the plasmid. Last step, motile bacteria swim towards the Target, where the plasmid is retrieved through sequencing processes.}
\label{fig:System_architecture}
\end{figure}

Figure~\ref{fig:System_architecture} describes the overall system architecture. The digital information are first encoded into nucleotides, and for this we analyse two different encoding techniques, described in Section \ref{sec:our_encoding}. The synthesized genes of the encoded nucleotides are then inserted into plasmids, and plasmids are taken up by bacteria through the process of transformation. To guarantee that such bacteria do not mobilize, we position them on solid agar (refer to Section~\ref{sec:wetexp} for further details about this process). We place these motility-restricted bacteria with at specific regions of the grid, as illustrated in Figure~\ref{fig:System_architecture}. To achieve our storage system, we position bacteria that contain the same chunk of information at the same spot in the grid. Therefore, each region of the grid identifies a cluster of bacteria that contain the same information, and every cluster stores different data from the others.

In the event that a read operation needs to be performed, motile bacteria are released from \emph{A}, swim towards the compartment, and then conjugate with the motile-restricted bacteria to retrieve the plasmids with the encoded information. Once this is complete, bacteria swim towards position \emph{C} to deliver the plasmids. Conjugation is a process where bacteria come together and form a physical connection that allows them to transfer plasmids between each other, and this process has a probability associated with it. At position \emph{C}, the plasmids are retrieved and sequenced to obtain the data and decoded back into digital format. 

A key requirement for the motile bacteria is the capability to swim towards an accurate point to conjugate with the motile-restricted bacteria, in order to retrieve the right plasmid with the encoded information,. This is where our proposed Molecular Positioning System (MPS) plays a role. As suggested by Okaie et al. \cite{Okaie}, it is possible to take advantage of engineered-bacterial chemotaxis in order for them to move towards a certain region, within the field of the chemoattractants, and this pairs well with our MPS proposal which is based on the receptor saturation addressing technique proposed by Moore and Nakano~\cite{Moore11}~\cite{Moore13}.



\begin{figure}[ht]
\center
\includegraphics[trim=0mm 0mm 0mm 0mm,clip,width=0.8\textwidth]{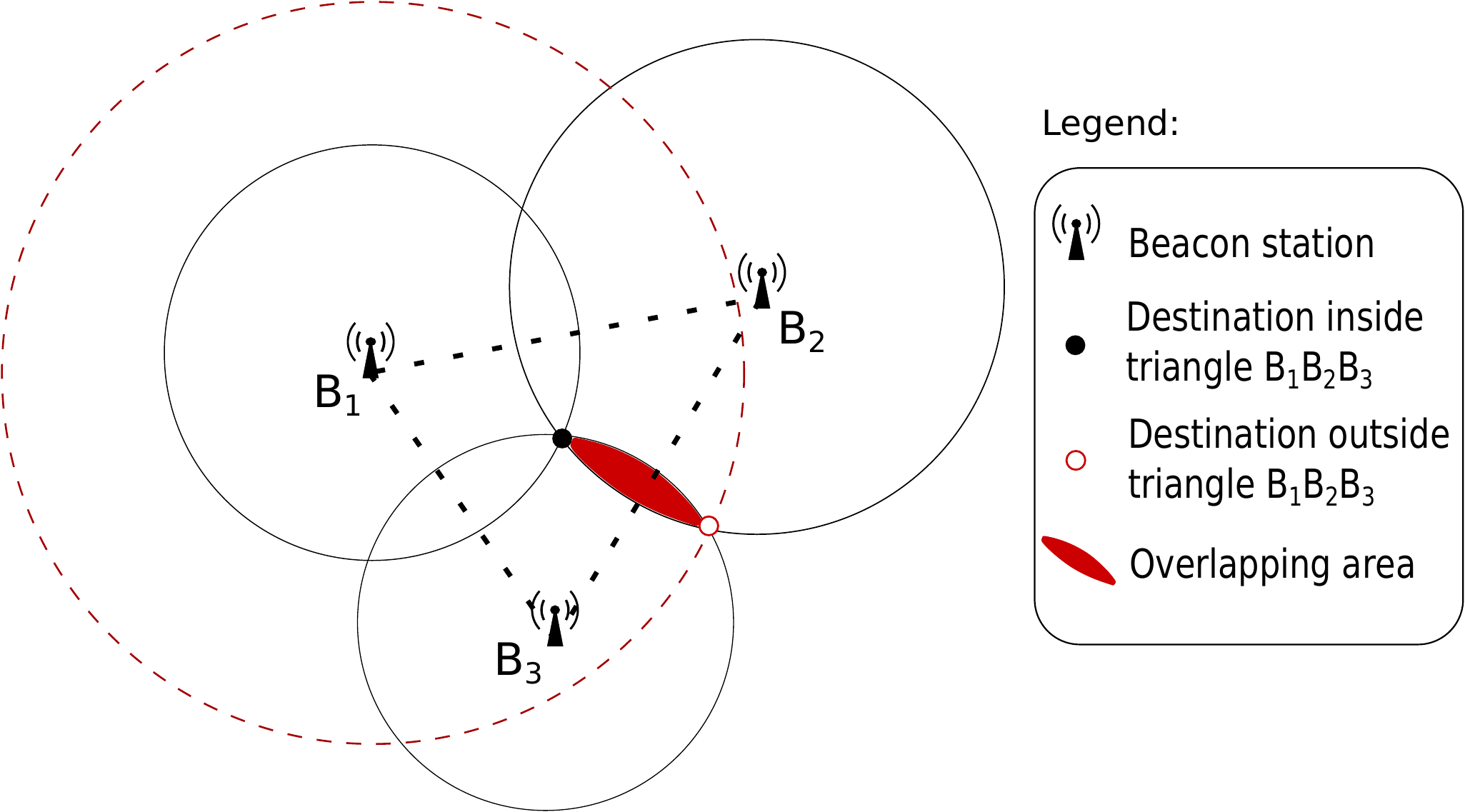}
\caption{An example of a general trilateration process. Drawing three circles that intersect at a point outside the $B{_1}B{_2}B{_3}$ convex hull, inevitably creates a red overlapping area, as shown here with the red circle centred in $B{_1}$ and the two black circles centred in $B{_2}$ and $B{_3}$.}
\label{fig:trilateration}
\end{figure}

While trilaterating an object is a difficult task, it is even more challenging when applied to chemical signalling environments. As shown in Figure~\ref{fig:trilateration}, it is always possible to draw three circles (each one centred on a beacon) that intersect at a specific point inside the convex hull formed by the beacons (i.e., the $B{_1}, B{_2}, B{_3}$ triangle), without creating an overlapping area. On the contrary, it is not possible to intersect at a point outside the convex hull avoiding an overlapping area, and the farther the intersection point sits from the convex hull, the bigger the resulting overlap is.

This geometric fact is particularly important for our proposed approach, since it entails that the achievable precision is inversely proportional to the size of the overlapping area. Indeed, in our proposed solution the motile bacteria react chemotactically towards their destination, until the receptors for each of the three chemoemissions are saturated. Once that the saturation has been achieved for all receptors, the motile bacteria perform random step-walks until they eventually drift too far away from the desired location. 

\subsection{Chemoreception and Mobility Model}
\label{subsubsec:behaviour}

In our work, we use the mobility model proposed for engineered bacteria by Okaie et al.~\cite{Okaie}. At each time step, which occurs every $\Delta t$ seconds, a bacterium senses the concentration of chemoattractants emitted from the beacons. We assume the concentration of chemoattractants is approximated as exponentially decreasing with respect to the square distance from the beacons. 

The concentration of attractants sensed at current position $(x,y)$, namely $C_{A}(x,y)$, is taken into account in the next time step only if it is lower than the value expected at the destination, namely $C_{A}(x_{d},y_{d})$. This criterion enables a bacterium to swim away from a beacon if it gets too close, with respect to its programmed destination, and this behaviour is represented as follows:
\begin{equation}
  \label{concentration}
    C_{A}(x,y)=
    \begin{cases}
      e^{\left(-d_{B}\left(x,y\right)^{2}\right)}, & if\ < C_{A}(x_{d},y_{d}) \\
      0, & otherwise.
    \end{cases}
\end{equation}


The engineered motile bacteria senses the chemoattractants concentrations within range $D_{A}=[-\psi_{A}, \psi_{A}]$ and select the angle $\Psi_{A}$ which results from the highest summation of $C_{A}$:
\begin{equation}
\label{Psi_A}
\Psi_{A}=\max\limits_{\psi \in D_{A}} \sum\limits_{i=1}^{m} C_{A_{i}}(x_{\psi}, y_{\psi}),
\end{equation}
where \textit{m} is the number of \textit{beacons}.

A random component $\Phi$ is necessary to emulate the typical run-and-tumble behaviour of the \emph{E. Coli} bacteria, which swim in a straight line at speed $v$, and correct their direction at specific intervals. Therefore, the random addend is:
\begin{equation}
\label{Phi}
\Phi = \pm \sqrt{D2\Delta t},
\end{equation}
where $\Phi$ is randomly chosen and $D$ is the rotational diffusion coefficient.

Finally, chemoattractants induced angle and random angle are summed up to compute a total drifting angle $\theta$:
\begin{equation}
\label{tot_drifting}
\theta_{t+\Delta t} = \theta_{t} + \Psi_{A} + \Phi .
\end{equation}

Substituting $\Phi$ from Equation\ref{Phi} we obtain:

\begin{equation}
\theta_{t+\Delta t} = \theta_{t} + \Psi_{A} \pm \sqrt{D2\Delta t} .
\end{equation}

\subsection{MPS - Molecular Positioning System}
\label{subsubsec:mps}
Our MPS aims to enable the engineered motile bacteria to approach (and remain within close vicinity of) a chosen position, leveraging the concentration of chemoattractants emitted by the beacons. We assume that the beacons are anchored at a known fixed position, and emit three discernible chemoattractants at a constant rate. The engineered motile bacteria are then programmed to move, targeting a specific molecular concentration for each of the three beacons, which means that matching the three different concentrations results in reaching the specific location desired. When a motile bacterium is released into the environment, the chemoattractant concentrations emitted by the beacons are used in two different ways: (i) to derive its relative position, and (ii) as a tool to move towards its target destination. Therefore, if a motile bacterium drifts too far away from a beacon, it is able to vary its own chemotactic behaviour and correct its position by following the specific chemoattractants.

In our proposal, we assume that, (i) the engineered motile bacteria know \textit{a priori} the three beacons locations, (ii) each beacon emits a unique chemoattractant, to allow the motile bacteria to infer how far each beacon is, and (iii) the beacons do not move around the environment. 


\section{DNA encoding}\label{sec:our_encoding}


In this section, we describe how digital data is encoded into DNA, as well as the process to package the data and the addressing. There are two basic operations to store and retrieve the encoded information, which could be a file or portion of a file:

\begin{itemize}
\item \texttt{Store (namespace(B1,B2,B3), filename$_i$)}: Stores the file into the location that is represented with concentration of (B1,B2,B3);
\item \texttt{Retrieve (namespace(B1,B2,B3), filename$_i$)}: This operation requires that bacteria are encoded with certain concentration of receptors on their surface, namely (C1,C2,C3). Engineered bacteria tend to drift towards the chemoemissions, until the receptors are saturated for each of the chemoemitter, therefore by this feature we can mobilize them towards specific points.
\end{itemize}

\begin{figure*}[ht]
\center
\includegraphics[trim=0mm 0mm 0mm 0mm,clip,width=1\textwidth]{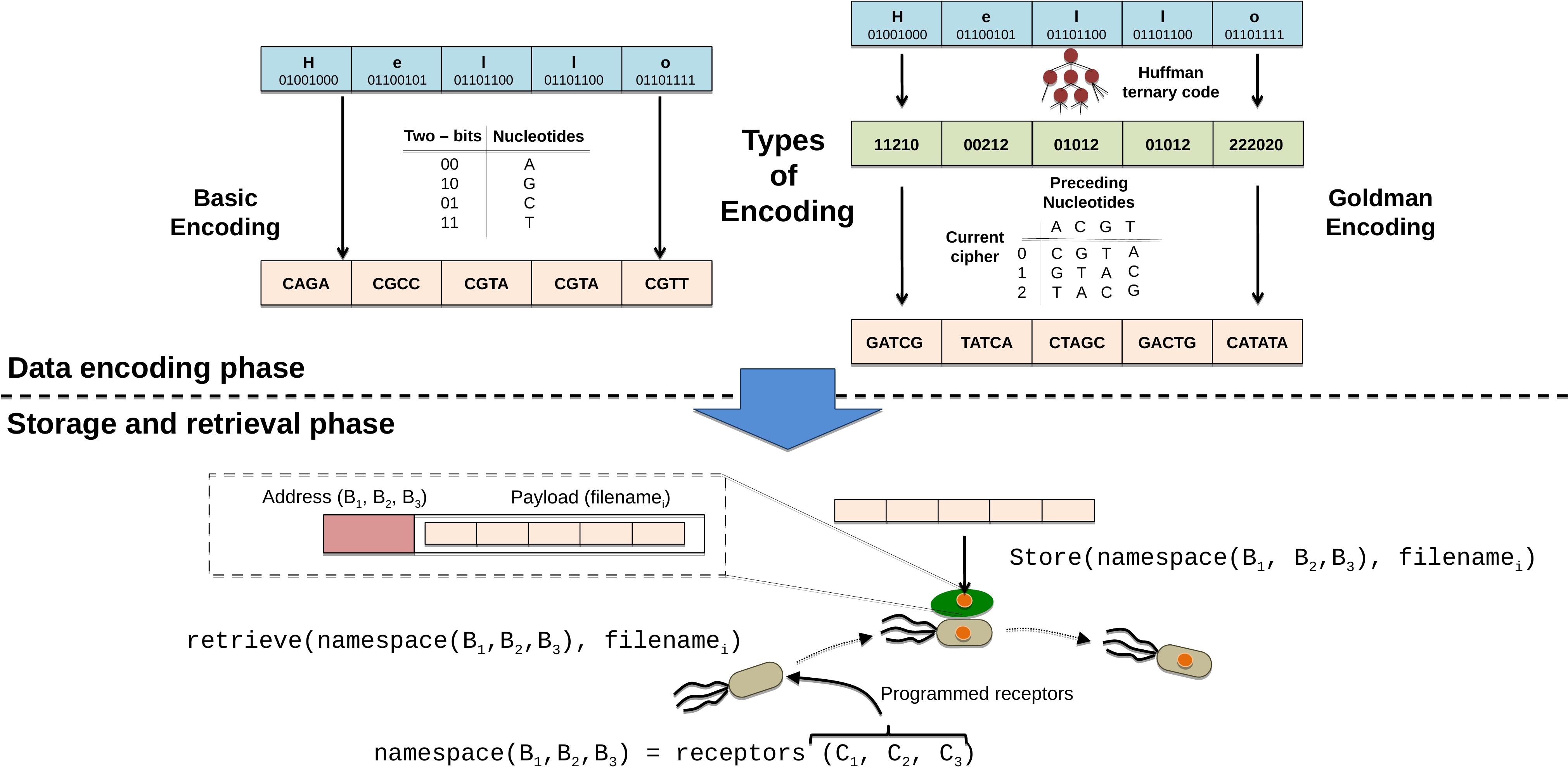}
\caption{The process demonstrating the types of encoding (Basic and Goldman encoding), and the production of digitally encoded DNA (payload), which is encapsulated through a virtual addressing that represents the location of the cluster. The virtual addressing encapsulation represents the concentration of receptors (C1, C2, C3, which corresponds to the beacon chemoattractants B1, B2, and B3) on the surface of the motile bacteria, which allows the bacteria to mobilize towards the cluster.}
\label{fig:encod_flow_conventional}
\end{figure*}

The entire process of the retrieval is illustrated in Figure~\ref{fig:encod_flow_conventional}, where the algorithm is presented in Algorithm~\ref{alg:conjugation_algorithm}. The end process of the encoding results in the payload DNA that contains the information, and this is encapsulated through a virtual addressing. The virtual addressing represents the location of the cluster of non-motile bacteria, which contain the information stored through the plasmid. The virtual addressing allows the bacteria to accurately swim directly to the cluster to conjugate with the non-motile bacteria to retrieve the plasmid with the encoded DNA.

\begin{algorithm}[t]
\KwIn{Positions of the clusters.}
\KwOut{Plasmids contained in the clusters.}
$t \leftarrow 0$, $\Delta t \leftarrow 50$\;
$conjugationTime \leftarrow 1500$, $timeLimit \leftarrow 7200$\;
\While{$t < timeLimit$ \textbf{or} \textit{data missing}}{
		\ForAll{$receivers$}{
			\textit{Move by one step}\;
			\If{$distance(retriever,cluster) < threshold$ \textbf{and} $shouldConjugate$}{
				$success \leftarrow tryConjugation()$\;
				\If{$success = True$}{
					\textit{Do not move for conjugationTime seconds}\;
					$shouldConjugate \leftarrow False$\;
				}
			}
		}
		$t \gets t + \Delta t$\;
}
\caption{Retrieving plasmids from the clusters}
\label{alg:conjugation_algorithm}
\end{algorithm}

We analysed two different techniques to convert binary data into nucleotides of the DNA. As described in~\cite{Microsoft}, there are plenty of encoding techniques used to transform digital data into strings of nucleotides. First, we describe the most simple conversion (2 bits-1 nucleotide match), that we call here \textit{basic encoding}, and the approach proposed by Goldman et al.~\cite{Goldman}, which improves over the basic encoding. Second, we describe how our DNA archive system can be utilized for content managing purposes, where clusters of motility-restricted bacteria are arranged based on the information priority. 


\subsection{Basic Encoding}\label{subsec:conventional_encoding}

As already mentioned, the Basic Encoding is based on a simple mapping. Since DNA is composed of 4 nucleotides (\textbf{A}denine, \textbf{C}ytosine, \textbf{G}uanine, \textbf{T}hymine; usually referred using the first letter). Using this technique we can encode $log_{2}(4) = log_{2}(2^{2}) = 2$ bits using a single nucleotide. In this way, we are able to use the 4 bases that compose the DNA strand to encode each byte of data. An example of the whole process is illustrated in Figure~\ref{fig:encod_flow_conventional}.

Even though this is a simple technique, it is not efficient and it is weak against DNA replication errors and mutations. As described in the next section, other proposed techniques~\cite{Goldman}~\cite{Microsoft} proved to be more efficient. 

\subsection{Goldman encoding}\label{subsec:goldman_encoding}

When we choose an algorithm to encode binary data into a sequence of nucleotides, we must keep in mind certain biological properties. For example, processes of DNA synthesis and sequencing are subject to a variety of errors, which include:

\begin{itemize}
\item \textbf{Insertion:} addition of one - or more - nucleotide to a DNA sequence;
\item \textbf{Deletion:} removal of at least one nucleotide from a DNA sequence;
\item \textbf{Substitutions:} mutation of one - or more - nucleotide. We can have two different types of mutation: \textit{transition}, which are interchanges of two-ring purines (i.e., adenine and guanine) or of one-ring pyrimidines (i.e., cytosine and thymine), or \textit{transversion}, interchanges of purine for pyrimidine bases which therefore involve exchange of one-ring and two-ring structures. These mutations can be differentiated in \textit{synonymous mutations}, which is when DNA mutation does not lead to an amino acid change, and \textit{non-synonymous mutations}, which can result in an amino acid change. This last kind of error can be split in two further categories: 
\begin{itemize}
\item \textbf{Missense mutation:} one amino acid replaced by another amino acid;
\item \textbf{Nonsense mutation:} amino acids replaced by a stop codon.
\end{itemize}
\end{itemize}

This nonsense mutation is the most problematic during translation (i.e., conversion of \emph{mRNA} into a protein) because it leads to truncated amino acid sequences, which in turn results in truncated proteins.
The probability of some forms of errors can be mitigated by encoding digital data into base 3, instead of using the technique illustrated in Section \ref{subsec:conventional_encoding}.
A base 3 encoding is proposed by Goldman \cite{Goldman}, which uses a rotational code to avoid \textit{homopolymers}, which are continuous repetitions of the same DNA base that can increase the likelihood of sequencing errors\cite{Barron}. The whole process is illustrated in Figure~\ref{fig:encod_flow_conventional}. 


First, binary data is encoded in base 3 using a Huffman code~\cite{Huffman}, which compresses data based on the frequencies of the characters. Since each character is composed of 8 bits, we need at least $2^{8} = 256$ base 2 representations. The closer we can get using base 3, results in $3^{6} = 729$ different ternary strings (with $3^{5} = 243$ representation we would not be able to convert all the characters). Thus, we are wasting $726 - 256 = 473$ states that are never used because for sequences of length five, we result in 13 characters missing. However, using the Huffman code, we can use five digits to represent the most common characters and six for the least common ones. In this way, we can maintain a reasonable overhead (i.e., one more digit) over a whole binary file.
Finally, the ternary data representation in converted into a nucleotide sequence using a rotational code 
to avoid continuous sequences of the same DNA base.

One may notice that, in the end, this encoding is less efficient than the basic encoding approach. In fact, to encode one byte the most obvious encoding requires 4 nucleotides, while the one proposed in~\cite{Goldman} uses 5-6 nucleotides (depending on the frequency of the character). On the other hand, we already pointed out that this last method is less error prone, and thus more reliable. On the other hand, this technique increases significantly the whole strand length (i.e., the file size). A trade-off between the two alternatives might be the XOR-encoding proposed in~\cite{Microsoft}, which has a reliability similar to the one described by Goldman, but is twice as dense.

\subsection{Priority Content Management}\label{subsec:content_management}

Once the encoding process is achieved, the resulting data is placed into motile-restricted bacteria clusters, where each cluster represents different libraries of information. As previously stated, to achieve a reliable behaviour clusters should be placed within the convex hull defined by the three chemoemitters. Given that there is only a finite space available, this limits the number of clusters that can be placed within the convex hull.

\begin{figure}[ht]
\centering
\subfloat[Conventional placement.]{\includegraphics[trim=0mm 0mm 0mm 0mm,clip,width=0.48\textwidth]{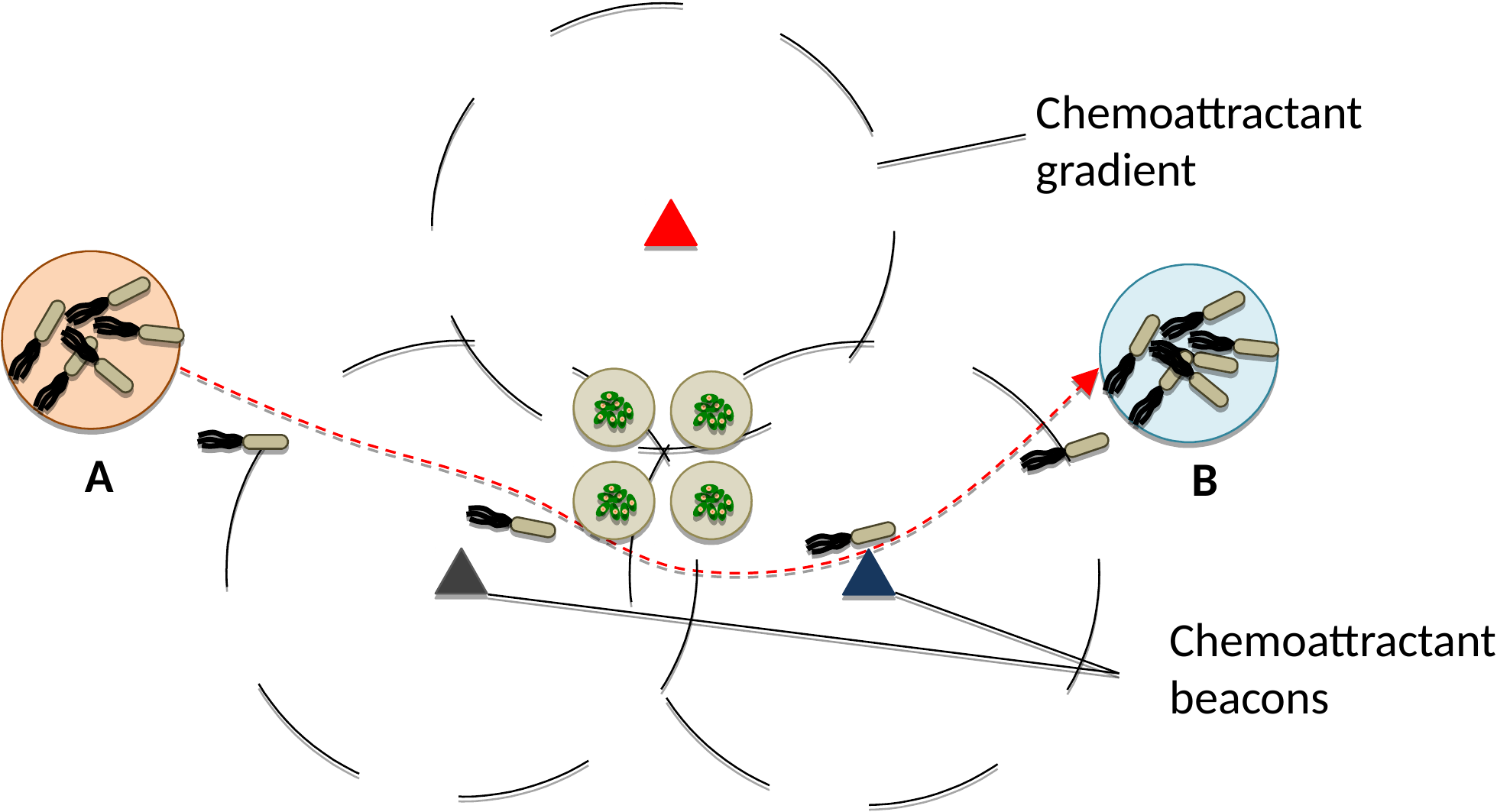}
\label{fig:comparision_data_management_a}}
\hfill
\subfloat[Prioritised content management placement.]{\includegraphics[trim=0mm 0mm 0mm 0mm,clip,width=0.48\textwidth]{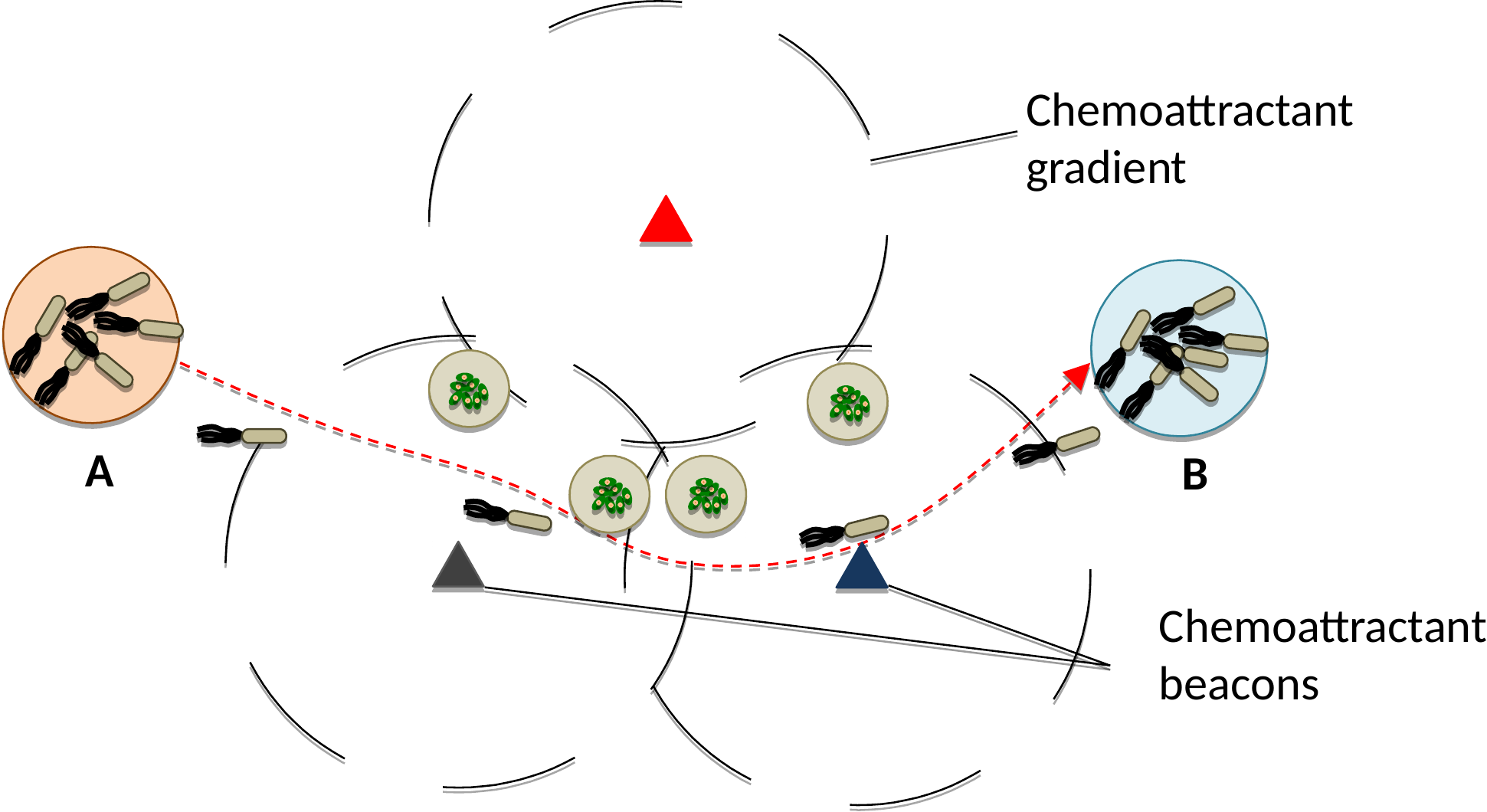}
\label{fig:comparision_data_management_b}}
\caption{Comparision of the clusters positioning.}\label{fig:comparision_data_management}
\end{figure}

In this section, we propose a content management system that takes advantage of this limitations and provides the opportunity to prioritize information. In particular, we propose a content system where higher priority information is stored into clusters placed inside the convex hull, whereas the lower priority information is stored outside the convex hull, in less reliable clusters.
The corresponding algorithm is shown in Algorithm~\ref{alg:content_distr}. As an example, which is illustrated in Figure~\ref{fig:comparision_data_management}, we concentrate on four clusters of bacteria. For this we distribute the four clusters in two macro areas: one inside the triangle and one outside of it. Consequently, we can distribute different kind of content (i.e., data with different access priority) into the different clusters. From the model proposed in Section~\ref{subsubsec:mps}, we expect to have a higher chance of retrieving encoded information from the cluster placed inside the convex hull (i.e., inside the triangle). Thus, data with lower access priority are inserted in the clusters placed outside the convex hull.

\begin{algorithm}[t]
\KwIn{Number of clusters with low/high priority.}
\KwOut{Clusters containing the encoded data.}
$N + M =$ \textit{number of clusters}\;
\textit{Set priority of $N$ clusters to low}\;
\textit{Set priority of $M$ clusters to high}\;

\While{\textit{inserted(data) $<$ amount(data)}}{
 $stored \leftarrow False$\;
 \ForAll{$clusters$}{
  	\If{$notFull(cluster)$ \textbf{and} $prority(data) = priority(cluster)$ \textbf{and} $stored = False$}{
  		\textit{Store data into the cluster}\;
  		$stored \leftarrow True$\;
  	}
 }
}
\caption{Content distribution based on priority}
\label{alg:content_distr}
\end{algorithm}

For example, we can imagine how a web browser can be stored using this mechanism. Firstly, we want to be sure that all the core functionalities (i.e., retrieving and presenting information from the Web) can be retrieved at each access to the storage system. Consequently, we place these features in the cluster inside the convex hull. Finally, we insert in the clusters outside the convex hull all the additions that are not fundamentals for the functioning of the browser (e.g., high-level CSS implementations, add-ons, plugins, etc.). This scenario is represented in Figure~\ref{fig:comparision_data_management_b}.

\section{Simulations}
\label{sec:simulation}

In this section, we discuss the variety of simulations that we run, aiming to establish the accuracy of our positioning system and the efficiency of the whole DNA archive system. 

First, we conduct some analysis over the Molecular Positioning System (MPS) which is essential for our proposal, By running these simulations, we assess the ideal chemoemitters configuration, in order to achieve the most accurate positioning possible for the motile bacteria. Second, after learning that the best configuration is achievable by positioning all the clusters within a convex hull area, we assess the capability of retrieving a chunk of data from such a configuration, by using the aforementioned MPS. Last, we assess the feasibility of a content management system, by storing low-priority information into 2 out of 4 clusters, and positioning these clusters outside the convex hull.

\subsection{Positioning System}
\begin{figure}[ht]
\center
\includegraphics[trim=9mm 8mm 18mm 8mm,clip,width=0.6\textwidth]{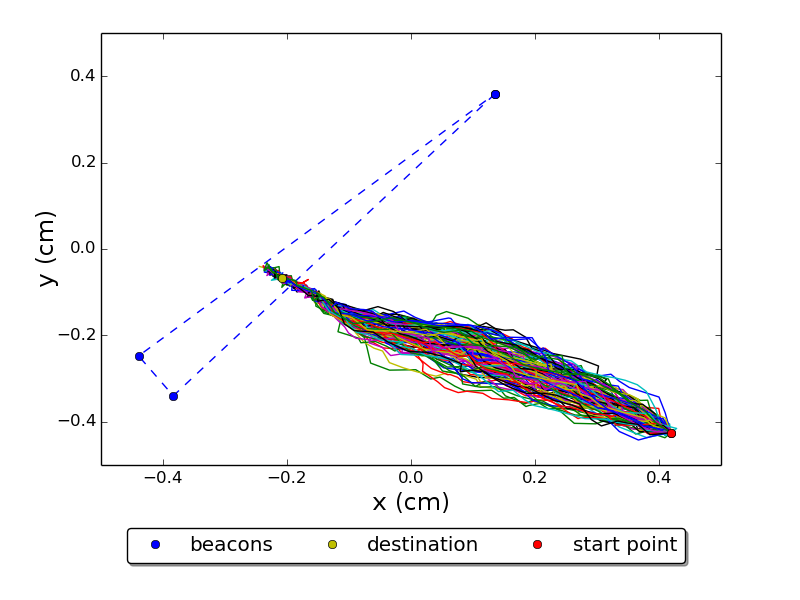}
\caption{Example of a simulation run. The engineered motile bacteria do not follow the same path, due to the random component in their movement algorithm~\cite{Okaie}.}
\label{fig:plot6}
\end{figure}

We execute a number of simulations to asses the efficacy of MPS; every independent simulation uses 100 engineered motile bacteria, where there are no interaction signalling between them. Table~\ref{tab:def_par} presents the default parameters used in our simulations.
Figure~\ref{fig:plot6} presents an example of a MPS positioning event. Initially, due to their random drifting angle, all the engineered motile bacteria take a different path as they mobilize towards the destination. However, as they approach the target, the paths tend to regroup and the engineered motile bacteria converge towards the target position.

\begin{table}[ht]
\renewcommand{\arraystretch}{1.0}
\caption{Default Parameters}
\label{tab:def_par}
\centering
\scalebox{1}{
\begin{tabular}{|l|l|}\hline
\bfseries Parameter & \bfseries Default Value\\ \hline
 $t$ & $1000~(s)$\\ \hline
 $\Delta t$ & $2\cdot 10^{-2}~(s)$\\ \hline
 $v$ & $5\cdot 10^{-3}~(cm/s)$\\ \hline
 $D$ & $5~(rad^{2}/s)$\\ \hline 
 $\psi_{A}$ & $3,49\cdot 10^{-2}~(rad)$\\ \hline
\end{tabular}
}
\end{table}

Our initial evaluation aims to determine the performance of the MPS based on variations of the target position. In particular, our goal is to compare the average positioning precision when the destination is placed inside and outside the convex hull. Furthermore, we want to verify if fluctuations appear in these two different cases.

\begin{figure}[ht]
\center
\includegraphics[trim=0mm 0mm 0mm 0mm,clip,width=0.6\textwidth]{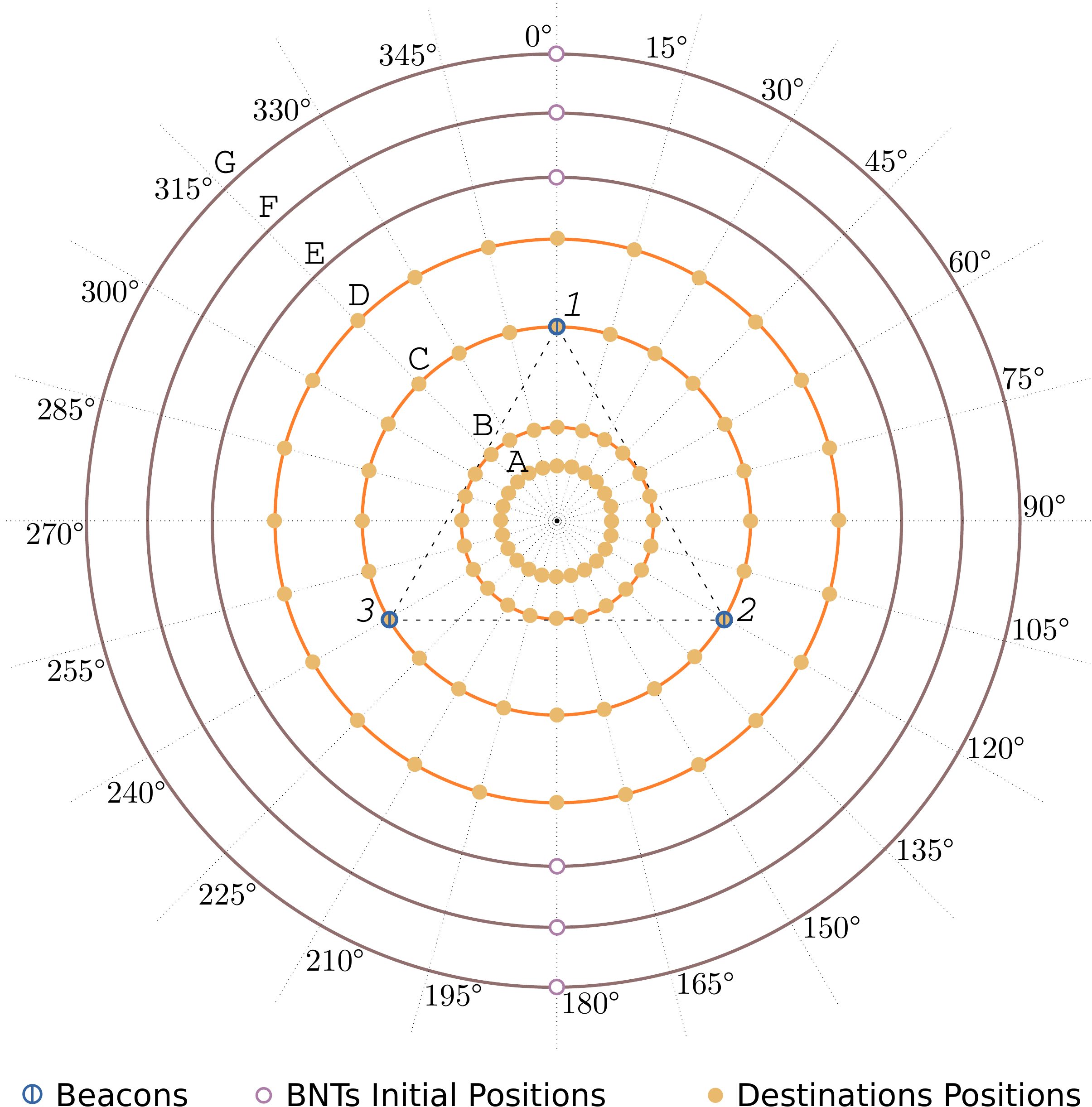}
\caption{Diagram that illustratess the performance inside and outside the beacons convex hull. For each initial bacterium position (spots on circles \textit{E}, \textit{F} and \textit{G}) we investigated every final destination (spots on circles \textit{A}, \textit{B}, \textit{C} and \textit{D}), which totals up to 576 independent simulations. The circles are centred at the triangle barycentre and, in order to meet spacial requirements, the radii sketched in this figure do not represent their real ratio with respect to the whole picture.}
\label{fig:circular_positioning}
\end{figure}

First, we position three beacons such that their convex hull is an equilateral triangle and place seven concentric circles of different radii, as shown in Figure~\ref{fig:circular_positioning}. For the four inner circles, we place the destination targets on the circle radii of $0.030~(cm)$, $0.058~(cm)$, $0.087~(cm)$ and $0.200~(cm)$, while the motile bacteria starting points are placed on the three outer circles, which radii are $0.300~(cm)$, $0.350~(cm)$ and $0.450~(cm)$. Each independent run consists of 100 engineered motile bacteria that share starting points and destination points, so that all the combinations of points are equally and separately evaluated. This results in a total of 576 independent simulations.

\begin{figure*}[ht]
\center
\includegraphics[trim=0mm 0mm 0mm 0mm,clip,width=1\textwidth]{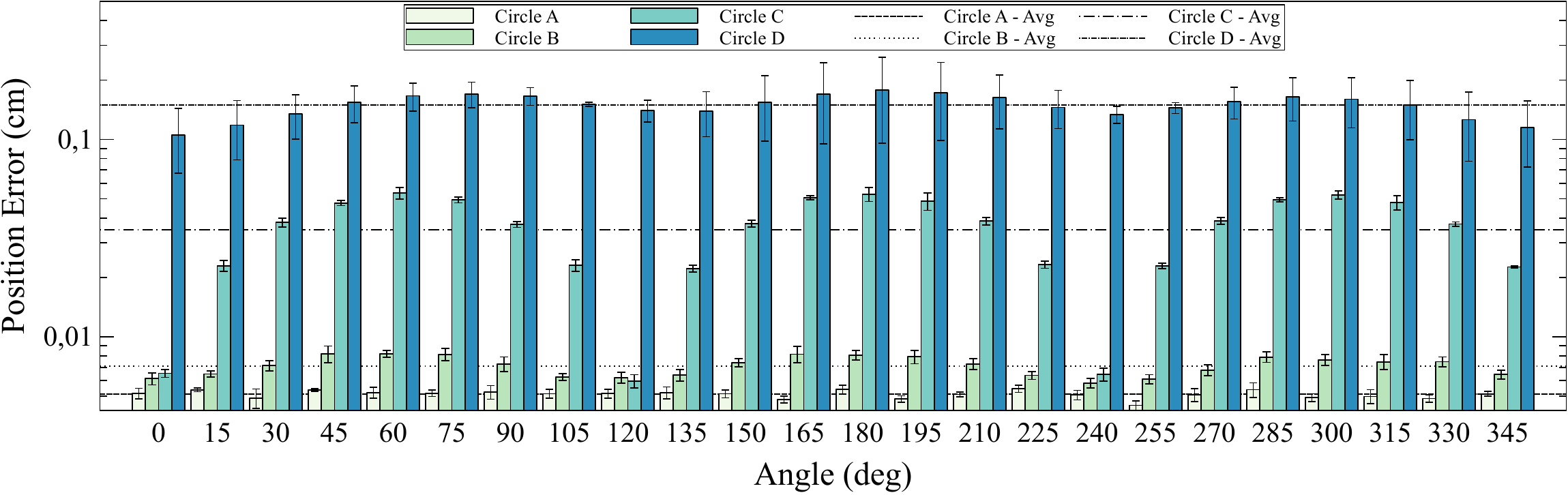}
\caption{Results are consistent and stable throughout all the simulations with destinations placed on \textit{Circle~A} and \textit{Circle~B}, whereas results clearly worsen when destinations are placed outside the convex hull, on \textit{Circle~C} and \textit{Circle~D}. We have chosen a logarithmic scale on Y axis to better depict such high variety of data. Each bar represents a set of 6 independent simulation, which totals up to 576 independent simulations.}
\label{fig:all_circles}
\end{figure*}

Figure~\ref{fig:all_circles} shows that the choice of the destination has a high impact on the results. We can observe that if the target location is contained within the convex hull, the results are consistent throughout the simulations: \textit{Circle~A} and \textit{Circle~B} bars highlight that the average positioning error remain under  $0,05~(cm)$ with a very small standard deviation.
On the other hand, \textit{Circle~C} and \textit{Circle~D} bars, which represent the destinations outside the convex hull, show that the positioning error grows considerably as the distance from the barycentre increases. As well as the average error, the standard deviation also increases and becomes more evident in \textit{Circle D} bars. These last results are particularly relevant, since they confirm the geometric concept that we expressed in Section~\ref{sec:architecture}, where the farther the destination point is from the convex hull, the more overlapping area of the three chemoattractants, which entails a lower precision.



\subsection{Retrieving Archived Information}\label{subsec:retrieving_archived_info}

After evaluating the feasibility of our MPS, we can use it to implement our \texttt{Reading} process from the DNA archive system, as described in Section~\ref{sec:architecture}. In order to prove the robustness of our model, we want to check if it is possible to retrieve the desired encoded file as we vary certain parameters in the simulator. Our simulations focus on three main points: the \textit{number} of engineered motile bacteria necessary to retrieve the whole file in a limited time slot, \textit{randomness} of engineered motile bacteria movement (related to factor $D$ described in Section~\ref{subsubsec:behaviour}), and the \textit{time} required to retrieve the whole file by varying the first two parameters. 

In order to do so, we create a file of $18.4~(KB)$ and encode it into DNA, which is distributed in the bacteria placed into four different clusters (e.g., encoding of the file generates $m$ plasmids, which are distributed in $n$ clusters that consequently contains $\frac{m}{n}$ plasmids). Each bacteria has an average capacity of 100 plasmids (i.e., normal distribution with average of 100 and standard deviation of 10), where each of the plasmids is composed of 200 base pairs (bps). We impose a limit of 120~(min) to retrieve and deliver the whole file.. We also assume the probability associated to conjugation as a normal distribution $N(0,1)$, and we use a threshold $(0.5)$ to decide the occurrence of the event. In order to start conjugation, we suppose that two bacteria should have a distance less than $10~(nm)$. To improve the likelihood of retrieving the data, we vary the number of bacteria - both retrievers for each cluster and bacteria used as storage inside the clusters - in the range $[10,150]$ with a step of 10 (e.g., 10, 20, 30). In this way, we keep a hypothetical ratio of 1 between the number of engineered bacteria used as retrievers and the number of the motility-restricted bacteria used as DNA storage. However, due to the random spatial distribution over the cluster area, we can not be certain that all bacteria are in a position that can be reached by the retrievers. To emulate this uncertainty, we impose as 50 the maximum number of bacteria that can simultaneously conjugate inside each cluster.

The experiment is based on three main locations: the start point (\emph{A}), the center of the triangle composed by the chemoemitters indicating the storage area (\emph{B}) and the center of the triangle composed of chemoemitters that indicate the end point (\emph{C}). These three points are vertically aligned; the distance between points \emph{A}-\emph{B} and \emph{B}-\emph{C} is $0.4~(cm)$. The clusters are placed at equal distance from point \emph{B} ($0.02~(cm)$ on both \emph{X} and \emph{Y} axis), composing a square. We place in point \emph{A} the engineered bacteria used as retrievers, switching on their chemoattractants receptors for the chemoemitters around the storage area. In this way, bacteria are able to swim towards the clusters to conjugate with the motility-restricted bacteria inside. Once conjugation occurs, two bacteria take about 120~(min) to exchange genetic material. Once horizontal gene transfer has happened, we switch off the chemoreceptors for the storage area and we switch on the ones that enable the engineered bacteria to reach point \emph{C}. Finally, when the bacteria reach their destination, they random-walk. In case that bacteria drift too much away from their destination, the receptors would be no more saturated and bacteria would try to approach the point again, as instructed by our MPS.

The evaluation is conducted by varying the following process and parameters of the simulator:

\begin{itemize}
\item \emph{Engineered motile bacteria}: 10 to 150 with a step of 10, creating 15 different simulations;
\item \emph{Random motility factor}: from 5 to 32 with a step of 3, leading to $10 \cdot 15 = 150$ different runs;
\item \emph{Encoding algorithm}: Basic and Goldman encoding.
\end{itemize}

All these different simulations are executed 10 times. Consequently, the results discussed in this section are calculated at an average of 1500 different simulations for each encoding algorithm (3000 simulations in total). 

\begin{figure}[ht]
\centering
\subfloat[Basic encoding.]{\includegraphics[trim=0mm 0mm 0mm 0mm,clip,width=0.48\textwidth]{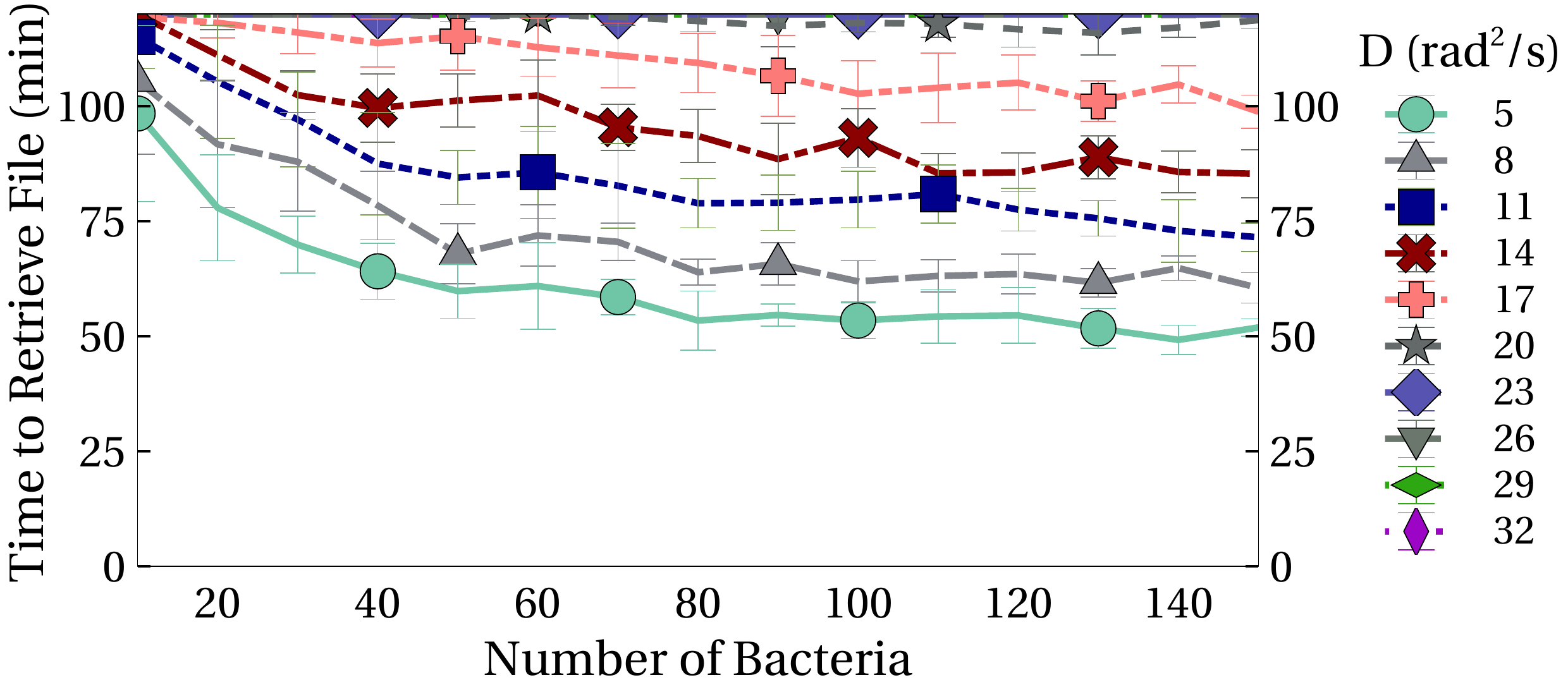}
\label{fig:time_basicEncoding}}
\hfill
\subfloat[Goldman encoding.]{\includegraphics[trim=0mm 0mm 0mm 0mm,clip,width=0.48\textwidth]{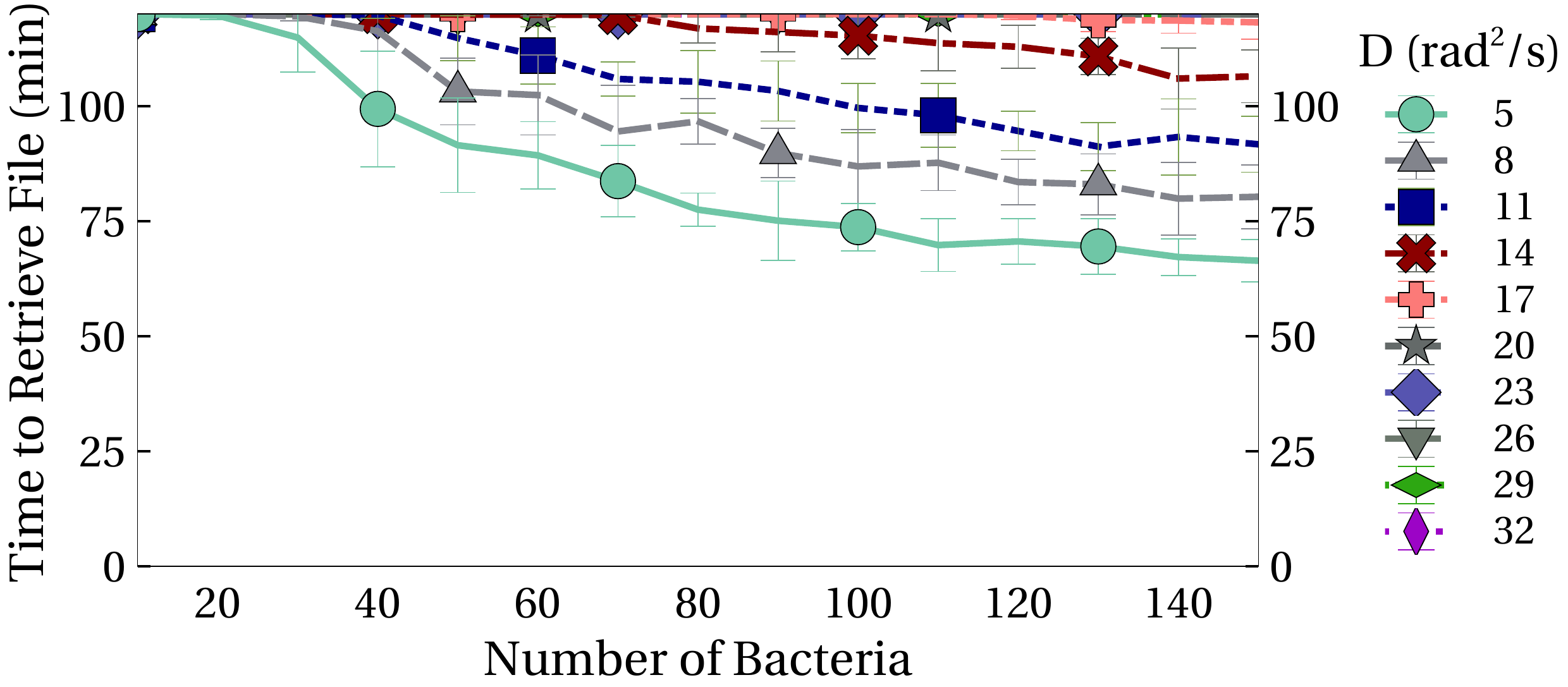}
\label{fig:time_GoldmanEncoding}}
\caption{Relationship between number of engineered motile bacteria, random factor $D$ and time needed to retrieve the whole file, both with the Basic and the Goldman encoding. If the time is greater or equal to 120 minutes, then the engineered motile bacteria were not able to retrieve all the information.}
\end{figure}
\begin{figure}[ht]
\centering
\subfloat[Basic encoding.]{\includegraphics[trim=0mm 0mm 0mm 0mm,clip,width=0.48\textwidth]{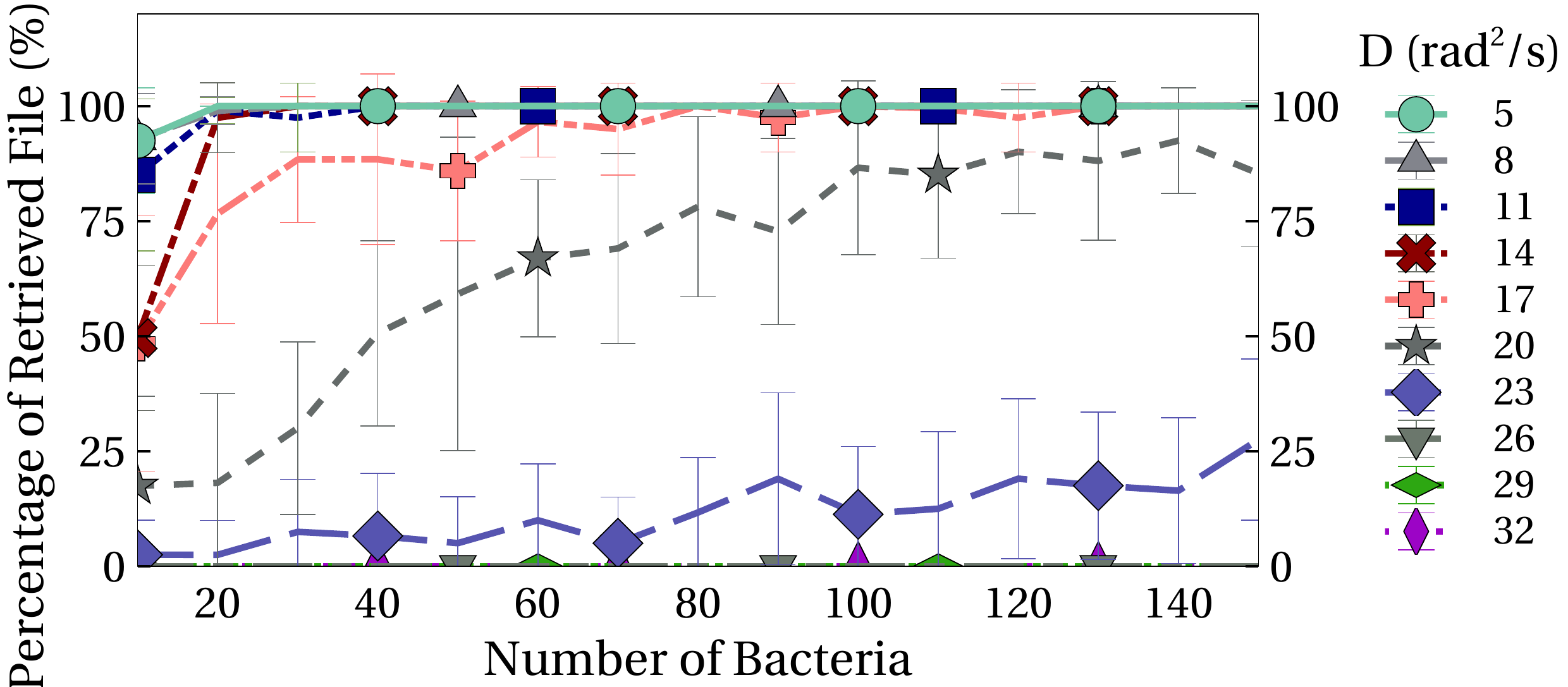}
\label{fig:percentage_basicEncoding}}
\hfill
\subfloat[Goldman encoding.]{\includegraphics[trim=0mm 0mm 0mm 0mm,clip,width=0.48\textwidth]{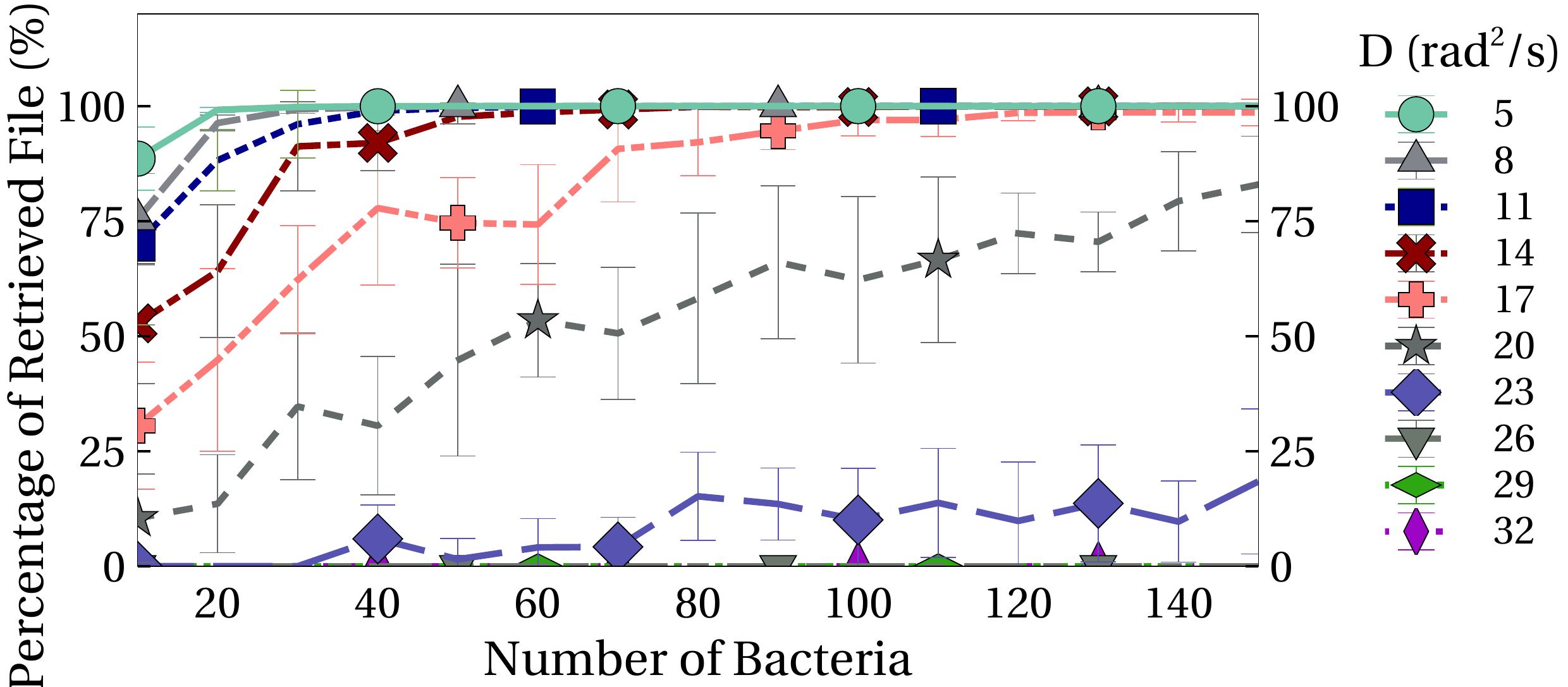}
\label{fig:percentage_GoldmanEncoding}}
\caption{Relationship between number of engineered motile bacteria, random factor $D$ and percentage of file retrieved, both with the Basic and the Goldman encoding. If the percentage is lower than 100\%, it means that the engineered motile bacteria were not able to retrieve the whole file in less than 120 minutes.}
\end{figure}

As we expected, Figure~\ref{fig:time_basicEncoding} and Figure~\ref{fig:time_GoldmanEncoding} shows that increasing the number of bacteria has a beneficial effect for the basic and Goldman encoding techniques. Indeed, the two figures show that increasing the throughput of our system, by increasing the number of engineered bacteria, leads to a lower time to deliver the whole file to the destination point. Unsurprisingly, both figures highlight the detrimental effect of increasing the bacteria randomness $D$, with respect to the time that it takes to retrieve the whole file at the destination. The same conclusions can be drawn by comparing Figure~\ref{fig:percentage_basicEncoding} and Figure~\ref{fig:percentage_GoldmanEncoding}, where lower number of bacteria and a high randomness in their movement makes it harder to direct them to their destinations, therefore harder to retrieve the whole file.

Comparing Figure~\ref{fig:time_basicEncoding} with Figure~\ref{fig:time_GoldmanEncoding}, and Figure~\ref{fig:percentage_basicEncoding} with Figure~\ref{fig:percentage_GoldmanEncoding}, the basic encoding performs clearly better than the Goldman encoding. This result is easy to explain, in order to be more resilient to mutations, the Goldman encoding uses more nucleotides than the basic encoding, and this overhead increases the time needed to complete the conjugation process. 

Considering that the default value for the random component $D$ is $5~(rad^{2}/s)$, it is remarkable that with $14~rad^{2}/s$ our system is still able to retrieve the whole file within the set time threshold, if enough bacteria (i.e., more than 80) are used. As a matter of fact, Figure~\ref{fig:time_basicEncoding} and Figure~\ref{fig:percentage_basicEncoding} shows that the basic encoding is still functioning even with randomness of $17~(rad^{2}/s)$.

\subsection{Content Management}

In Section~\ref{subsec:content_management} we discussed about how different topologies of clusters can lead to a content management system for the DNA archive system. Based on the reliability of the MPS, this means placing the more frequently accessed data inside the convex hull and the less priority contents outside of the convex hull defined by the beacons. Here, we test the file retrieval ratio (in percentage) achieved by the bacteria and how the two different encoding techniques affect the whole system performance. We execute the same set of simulations described in Section \ref{subsec:retrieving_archived_info}, however, we place the two uppermost clusters outside the convex hull area and on the same abscissa of the previous simulations, but horizontally aligned to the two emitters composing the base of the triangular convex hull.







\begin{figure}[ht]
\centering
\subfloat[Basic encoding.]{\includegraphics[trim=0mm 0mm 0mm 0mm,clip,width=0.48\textwidth]{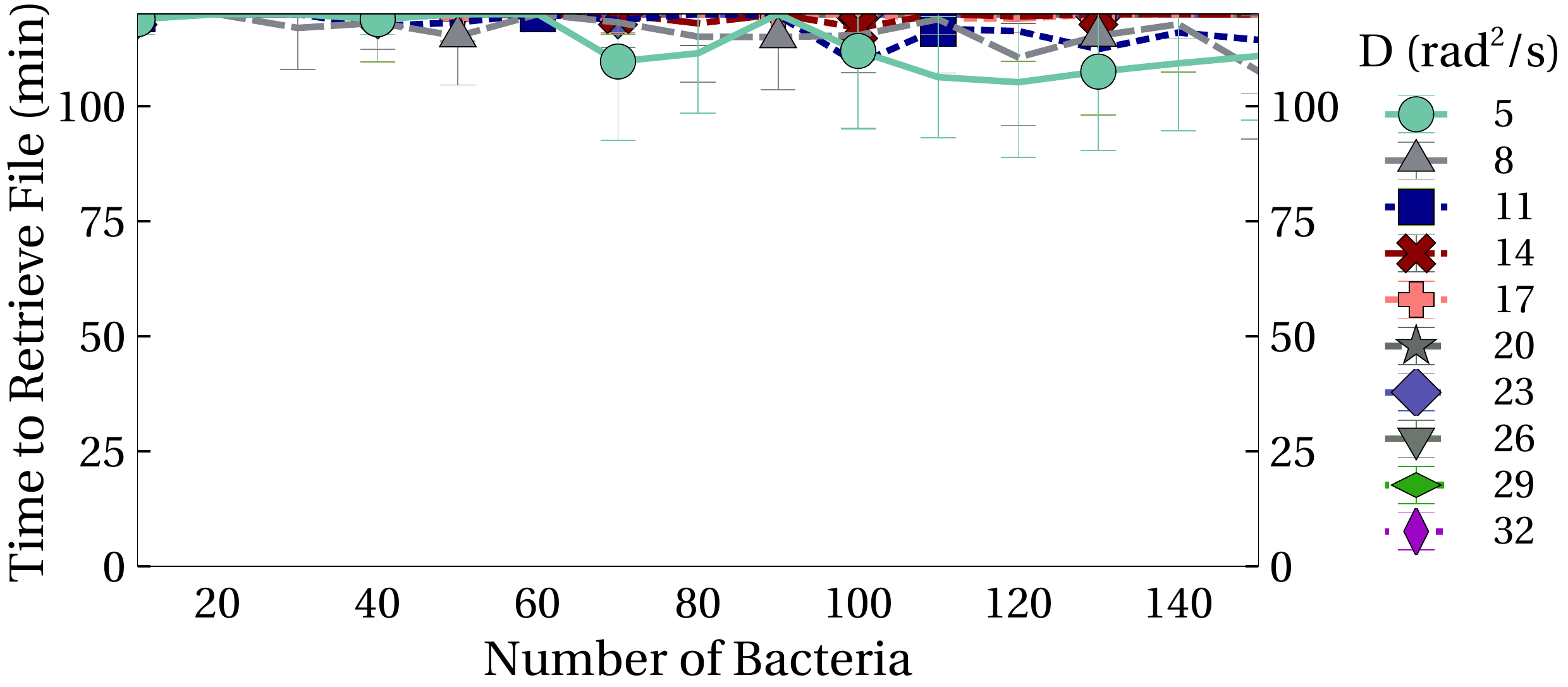}
\label{fig:time_2in_2out_basicEncoding}}
\hfill
\subfloat[Goldman encoding.]{\includegraphics[trim=0mm 0mm 0mm 0mm,clip,width=0.48\textwidth]{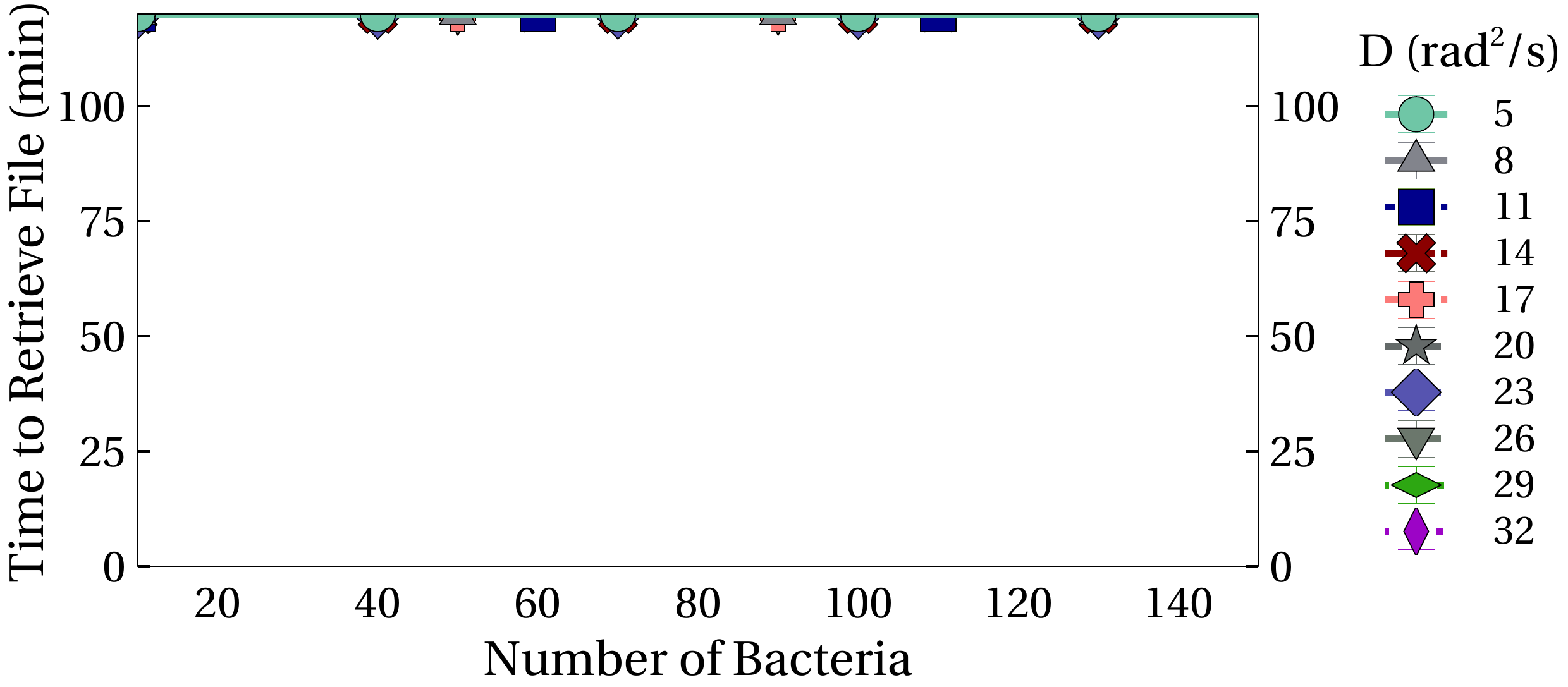}
\label{fig:time_2in_2out_GoldmanEncoding}}
\caption{Relationship between number of engineered motile bacteria, random factor $D$ and the time required to retrieve the whole file, both with the Basic and the Goldman encoding. If the time is greater or equal to 120 minutes, then the engineered motile bacteria were not able to retrieve all the information.}
\end{figure}

Analysing Figure~\ref{fig:time_2in_2out_basicEncoding} and Figure~\ref{fig:time_2in_2out_GoldmanEncoding}, it is easy to verify again that the Basic encoding performs better than the Goldman encoding technique even for the content management system. This is based on the results that shows that with sufficient bacteria and a reasonably low $D$ value, the content management system sometimes functions with the Basic encoding. However, the same does not hold for the Goldman encoding, which never enables us to retrieve the whole file within the $120~(min)$ threshold. 

\begin{figure}[ht]
\centering
\subfloat[Basic encoding.]{\includegraphics[trim=0mm 0mm 0mm 0mm,clip,width=0.48\textwidth]{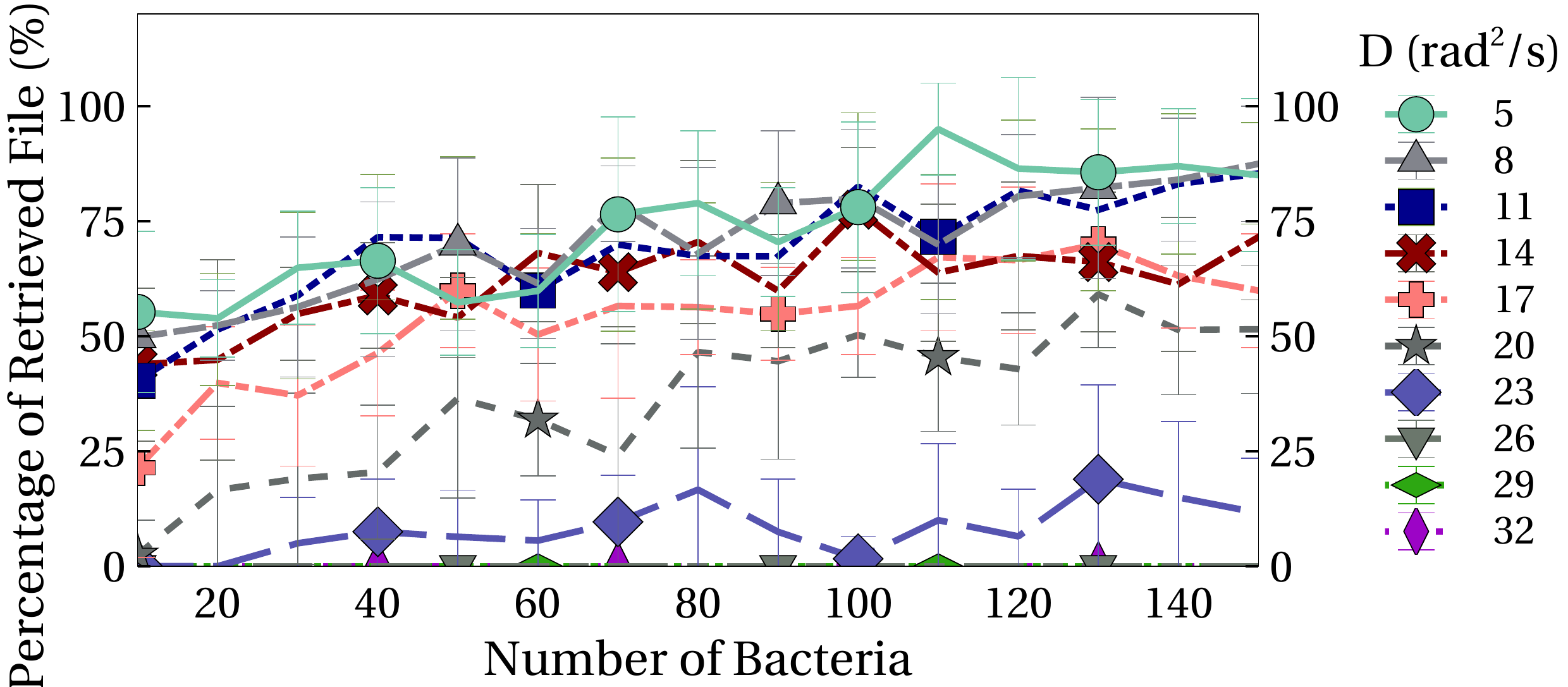}
\label{fig:percentage_2in_2out_basicEncoding}}
\hfill
\subfloat[Goldman encoding.]{\includegraphics[trim=0mm 0mm 0mm 0mm,clip,width=0.48\textwidth]{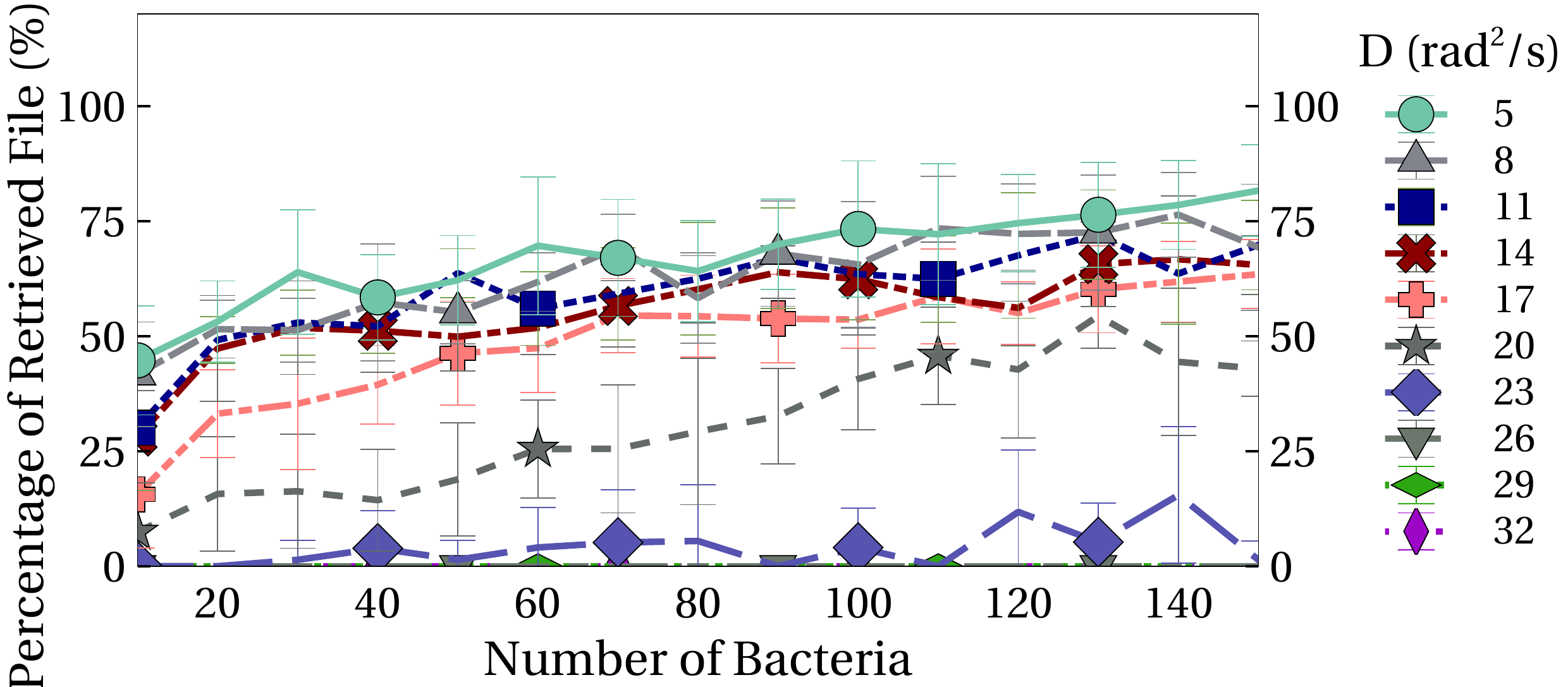}
\label{fig:percentage_2in_2out_GoldmanEncoding}}
\caption{Relationship between number of engineered motile bacteria, random factor $D$ and percentage of file retrieved, both with the Basic and the Goldman encoding. If the percentage is lower than 100\%, it means that the engineered motile bacteria were not able to retrieve the whole file in less than 120 minutes.}
\end{figure}

Looking at Figure~\ref{fig:percentage_2in_2out_basicEncoding} and Figure~\ref{fig:percentage_2in_2out_GoldmanEncoding}, we want to remark that each graph line is the resulting average of 10 different simulations. Therefore, we can conclude that on average we cannot be sure that the whole file can be retrieved, but the standard deviation bars for the Basic encoding shows that with sufficient bacteria and a low randomness, the successful retrieval can be achieved.

On the contrary, with the Goldman encoding the whole file is never retrieved since not even a single standard deviation bar hits the 100\% mark. However, with a content management system in mind, losing certain data is not inherently a critical issue, where this could be compensated by error detection and correction algorithms. These algorithms could be set at the destination point to assess the integrity of the file and, if possible, fix it. The other option is for the archive owner to decide to place in the outermost clusters the lower priority data, which can be retrieved without any time constraints or hard deadlines.



\section{Wet Lab Experiments}
\label{sec:wetexp}

\begin{figure}[t]
\center
\includegraphics[trim=0mm 0mm 0mm 0mm,clip,width=0.55\textwidth]{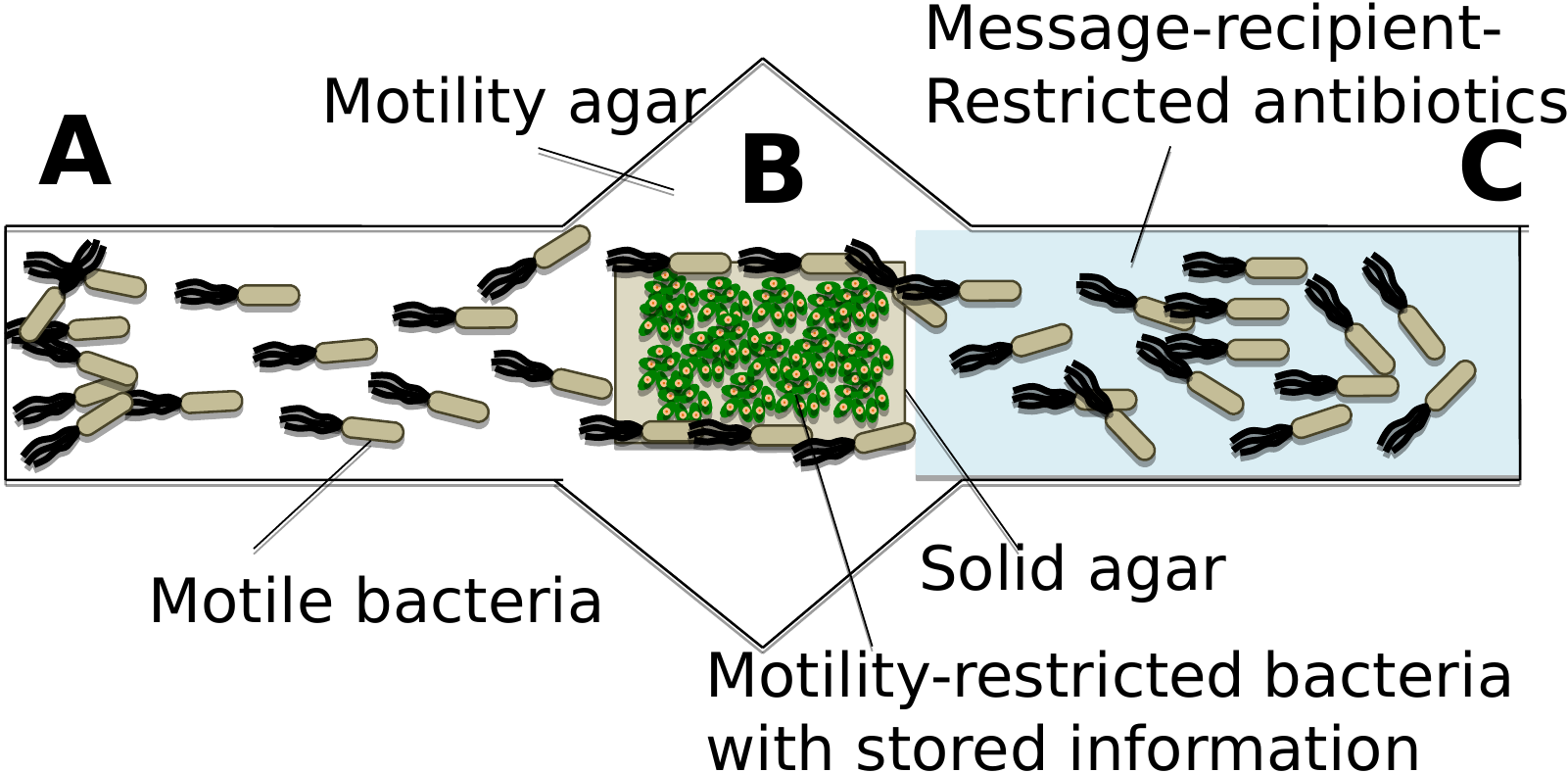}
\caption{Illustration of the wet lab experimental set up, where the motility-restricted bacteria with the encoded "Hello World" is placed at B. Non-motile bacteria swim from A towards C and pick up the message along the way.}
\label{fig:wetlab_illust}
\end{figure}

\begin{figure*}[ht]
\includegraphics[trim=0mm 0mm 0mm 0mm,clip,width=1.0\textwidth]{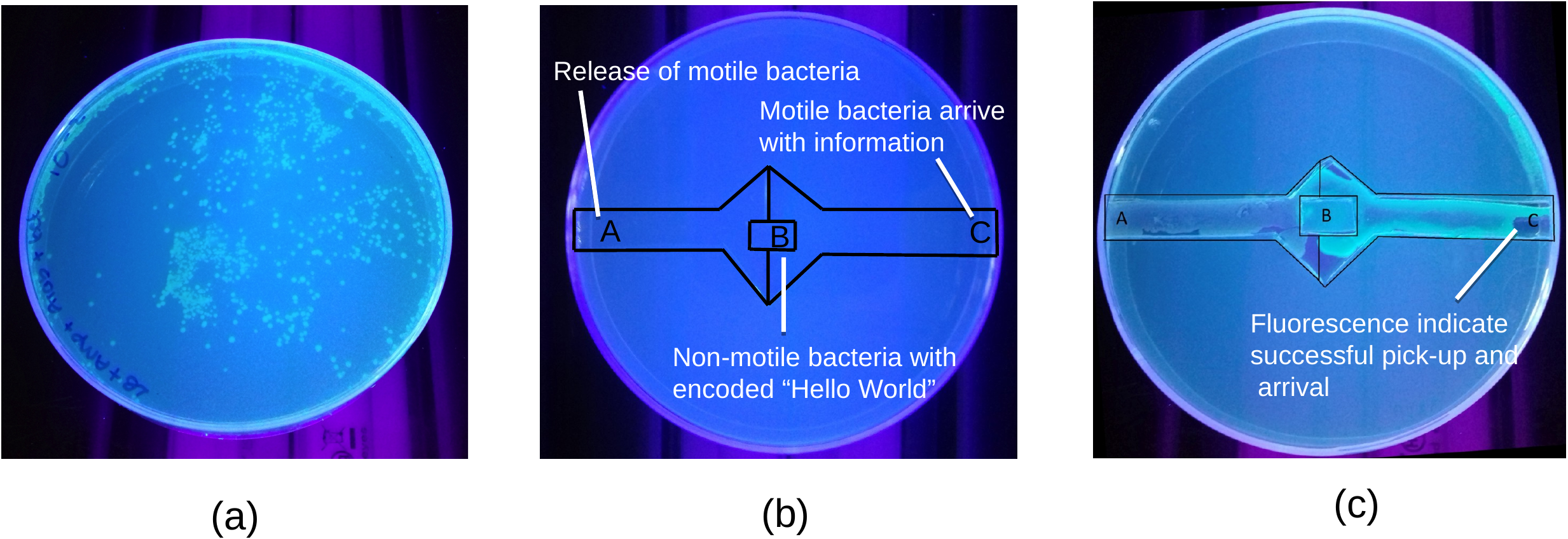}
\centering
\caption{Wet lab experimental result demonstrating the successful pick-up of the encoded "Hello World" message from the non-motile bacteria in B by the motile bacteria released at A. After picking up the message, the motile bacteria swims towards C to deliver the message. (a) demonstrates the successful conjugation process, (b) presents the engineered agar plate illustrating the placement of the non-motile bacteria with the encoded plasmid of "Hello World", and the channel for the motile bacteria to swim from A to C, (c) results of the experiment, where the fluorescence indicates the motile bacteria successfully conjugated with the non-motile bacteria to pick up the encoded plasmid with antibiotic resistance gene and surviving the antibiotics.}
\label{fig:Wet3}
\end{figure*}

In this section we discuss wet lab experiments to demonstrate our concept of \texttt{Reading} from DNA message-encoded plasmids that are stored in motility restricted bacteria. Figure~\ref{fig:wetlab_illust} illustrates our wet lab experimental setup that is based on engineering an agar plate to have a channel with motility agar. The surrounding portions of the agar plate, as well as the centre island (\emph{B}), is hard agar that ensures no motility can occur. In this experiment, the motile bacteria are released from \emph{A}, and swim towards \emph{C}, and along the way conjugate with the motility-restricted bacteria in \emph{B}. The conjugation process leads to the motile bacteria picking up the plasmid with the encoded information, which also contains the antibiotic resistance gene from the motility-restricted bacteria. The encoded information in the motility-restricted bacteria is {\bf Hello World}. The successful pick of the plasmid allows the motile bacteria to survive the antibiotics in \emph{C}, the destination of the message, as illustrated in Figure~\ref{fig:Wet3}b.

The first step of the experimental work, after the \emph{pWITGLO} plasmid (please see Appendix A for Material and Methods) was constructed, was to demonstrate the conjugation process, and this is illustrated in Figure~\ref{fig:Wet3} (a).  Two \emph{E.coli} K-12-derived commercially available recombinant cloning strains were used; \emph{Novablue} and \emph{HB101}. \emph{Novablue} is Tetracycline resistant, \emph{HB101} is \emph{Streptomycin} resistant. \emph{Oligonucleotide} primers were designed to amplify the full GFP gene from \emph{pGlo} using PCR, with the \emph{"Hello World"} message contained in the reverse primer immediately after the stop codon, resulting in the message immediately downstream of the gene as a single stranded overhang. \emph{Klenow} polymerase was then used to fill in this overhang, resulting in a complete double stranded PCR product; the GFP gene and the message. This "filled in" product was then prepared for ligation/insertion to a cloning vector(\emph{TOPO pEXP5/CT}) before being transformed into \emph{Novablue} cells. The resulting recombinant plasmid was named \emph{pWITGLO}. For the conjugation demonstration experiment, fresh growing cells of each strain were mixed for 1 hour to encourage conjugation and then spread onto agar containing the correct inducer \emph{IPTG} (induces GFP expression and fluorescence), \emph{streptomycin} (\emph{HB101} can tolerate this, \emph{Novablue} cannot) and ampicillin (selects for the plasmid). All resulting colonies growing and fluorescing are \emph{HB101} having taken up \emph{pWITGLO} from \emph{HB101}, now resistant to both antibiotics.  


The next experiment is to demonstrate the DNA encoded information Reading process. LB agar plates containing \emph{kanamycin}, so neither strain can grow, were used to flank motility channels. These motility channels consisted of motility agar, which is softer, allowing motility and the spread of cells. Figure~\ref{fig:Wet3} (c) shows, after inoculation, the successful Reading process. Zone \emph{A} contained \emph{Streptomycin} so \emph{HB101} (without \emph{pWITGLO}) can only  grow, and the plate was inoculated at \emph{A} with \emph{HB101}. This strain was seen to grow/move towards \emph{B} throughout the motility agar. In the middle of the plate, \emph{B}, was a square plug of motility agar containing \emph{IPTG}, \emph{tetracycline} and \emph{ampicillin} inoculated with \emph{Novablue} containing \emph{pWITGLO}, which was fluorescent. This strain was motility-restricted to zone \emph{B} as it is sensitive to the surrounding \emph{Streptomycin}. \emph{HB101} was unable to grow within zone \emph{B} due to the presence of \emph{Tetracycline}. Conjugation occurred at the interface of Zones \emph{A} and \emph{B}. In zone \emph{C} was motility agar with \emph{IPTG}, \emph{ampicillin} and \emph{streptomycin}. As illustrated in the picture, only \emph{HB101} cells that have moved from \emph{A} to \emph{B}, picked up \emph{pWITGLO} from \emph{Novablue} have grown here, now resistant to both antibiotics and able to fluoresce. This occurred within 72 hours, at a distance of 10 cm from \emph{A} to \emph{C}.  The fluorescence in \emph{HB101} is higher than \emph{Novablue}, with an increase of approximately 250 percent in fluorescence. \emph{HB101} cells with \emph{pWITGLO} were inoculated from \emph{C} to fresh liquid broth and \emph{pWITGLO} plasmid was purified from the resulting culture, with very high amounts of plasmid recovered; 380 \emph{ng/µl} concentration, 38 \emph{mg} of total plasmid DNA was recovered from 3 \emph{mls} of culture (10 \emph{mg} of wet cells). DNA sequencing of the GFP gene and DNA message confirmed transfer of the message to point C. The performance results are presented in Table~\ref{tab:perf_results}.

\begin{table}[ht]
\renewcommand{\arraystretch}{1.0}
\caption{Performance results}
\label{tab:perf_results}
\centering
\scalebox{1}{
\begin{tabular}{|p{5.5cm}|l|}\hline
\bfseries Factor & \bfseries Results (unit) \\ \hline
 Time taken for bacteria from \emph{A} to reach \emph{C} & $72~(hours)$\\ \hline
Relative fluorescence of \emph{HB101} cells with \emph{pWITGLO} versus \emph{Novablue} & $250~(percent)$\\ \hline
Concentration of \emph{pWITGLO} recovered from 10 \emph{mg} of wet cells  & $380~(ng/µl)$ \\ \hline
\end{tabular}
}
\end{table}

\section{Conclusion}
\label{sec:conclusion}

The continual growth of information created by the Internet has led to researchers searching new solutions for data storage. Digitially encoded DNA has recently been proposed as a promising solution for storing large quantity of data anticipated in the near future. In this paper, we combined the concept of digitally encoded DNA with molecular communications. The molecular communication system proposed is bacterial nanonetworks, where the bacteria mobilize and move towards spatial addresses based on engineered receptor saturation, in order to pick up the DNA encoded information from stationary non-motile bacteria. We term this directed movement as Molecular Positioning System, and is based on the concept of trilateration of a location placed inside and outside the convex hull of the chemoattractant field. Our positioning technique through simulations has shown that they can can consistently keep their positioning error under under $0,05~(cm)$ if their destination is placed within the convex hull. Through our simulations, this has shown to very beneficial, where we compared the realiability of the bacteria to pick up information from two types of encoding techniques, which includes Basic and Goldman encoding. Our simulations has shown that the basic encoding basically achieves higher performance compared to the Goldman, given the lower overhead in the size of the DNA after encoding. We have also presented the concept of prioritizing data placement in the non-motile clusters at various locations depending on the overlapping chemoattractants. This allows the numerous data to be placed in different regions, in order to cater for speed and reliability of retrieving the data in a timely manner. Lastly, the paper presented wet lab experimental work to demonstrate the validity of using non-motile bacteria to store a short message, where motile bacteria conjugate to pick up this message and deliver it to another location. Our work has demonstrated the feasibility of developing an archive system for digitally encoded DNA, which to date has remained as a big challenge for researchers.

\appendix
\section{Materials and Methods}\label{sec:appendixA}
\subsection{Construction of pWITGLO}
\emph{Oligonucleotides} were designed to amplify the complete GFP gene from \emph{pGlo} plasmid (\emph{BioRad}) via PCR, with the \emph{Hello World} message contained in the reverse primer immediately after the stop codon in a reverse complement orientation (in bold) (GFP F: \texttt{atggctagcaaaggagaaga}, GFPM R: \texttt{tgcgtacgctagaacgagggttctaacgtacgtacgggcg\\tctgttatttgtagagctcatcca}). \emph{Oligonucleotides} were supplied by Eurofins, MWG operon, Germany. PCR conditions used to amplify the GFP gene and message were as follows; each 15 $\mu$L PCR reaction mixture contained 7.5 $\mu$L Q5® High-Fidelity DNA Polymerase Master Mix (NEB), 15 \emph{pmol} of each primer and 15 \emph{ng} \emph{pGLO} plasmid DNA. PCR conditions: 1 cycle of 95 \textdegree{}C for 5 min, 30 cycles of 95 \textdegree{}C for 1 min, 56 \textdegree{}C for 1 min, 72 \textdegree{}C for 1 min, 1 final extension stage of 8 minutes. Negative controls contained water in place of DNA. PCR products were analysed by agarose gel electrophoresis. 

PCR products were cleaned using the Clean \& Concentrator\textsuperscript{TM} 5 kit (Zymo Research) as per manufacturer's instructions. DNA polymerase I, large Klenow fragment, (Thermofisher Scientific) was used to fill in the DNA message overhang on cleaned PCR product as per manufacturer's instructions. This double-stranded GFP-message construct was then subjected to A-overhang addition; this $\mu$L reaction contained \emph{Taq} DNA polymerase ((\emph{Taq polymerase Amresco})) and 10 \emph{mM} \emph{Datp} per reaction (\emph{Amresco}), to ready the DNA for ligation to a cloning vector. The double-stranded A-overhang GFP-message construct was ligated to \emph{pEXP5-CT/TOPO} vector (ThermoFisher Scientific) as per manufacturer's instructions. Plasmids were transformed into Novablue Gigasingles\textsuperscript{TM} (\emph{Novagen}) with positive clones identified by GFP-message PCR assay. Plasmid DNA (\emph{pWITGLO}) was purified using the Monarch® Plasmid Miniprep Kit (New England Biolabs). Sequencing of plasmid insert was carried out in triplicate using the vector-specific universal primers \emph{T7} promotor and \emph{T7} terminator, with all sequencing carried out by GATC Biotech, Germany. Nucleotide sequences were analysed using Blastn software from the NCBI database. 

\subsection{Bacterial strains}
\emph{E.coli} strains used were \emph{E.coli Novablue} (\emph{Novagen}) (genotype (\emph{endA1 hsdR17(rK12-mK12+}) \emph{supE44 thi-1 recA1 gyrA96 relA1 lac F'[proA+B+ laclqZ'M15::Tn10] (TetR))} and \emph{E.coli HB101} (BioRad) (genotype (\emph{F- Lambda- araC14 leuB6(Am) DE(gpt-proA)62 lacY1 glnX44(AS) galK2(Oc) recA13 rpsL20(strR) xylA5 mtl-1 thiE1 hsdS20(rB-, mB-)}. 

\subsection{Growth conditions}
\emph{E.coli} strains were grown on LB agar (Sigma Aldrich) with appropriate antibiotic; 12.5 $\mu$g/ml \emph{Tetracycline} and \emph{Streptomycin} 25 $\mu$g/ml (Sigma Aldrich) for \emph{Novablue} and \emph{HB101} respectively, at 37\textdegree{}C. \emph{Novablue} cells transformed with \emph{pWITGLO} were grown and maintained on LB media containing 12.5 $\mu$g/ml \emph{Tetracycline}, 100 $\mu$g/ml \emph{Ampicillin} (Sigma-Aldrich) and 0.5 \emph{mM} \emph{IPTG} (Sigma-Aldrich). 

\subsection{Motility and conjugation experiments}
Motility experiments were carried out by stab-inoculating motility agar (Sigma Aldrich) and incubating at 37\textdegree{}C for 3 days, observing motility. Conjugation experiments were carried out by harvesting Novablue containing \emph{pWITGLO} and \emph{HB101} cells from mid-log phase cultures (grown in LB broth and appropriate antibiotic), washing cells three times in 0.85 percent saline solution, resuspending both cell cultures together in LB broth and incubating with gentle agitation at 37 \textdegree{}C for one hour, before plating onto the surface on LB agar containing \emph{Streptomycin} 25 $\mu$g/ml, 100 $\mu$g/ml Ampicillin and 0.5 \emph{mM} \emph{IPTG}.

\subsection{DNA Message Transfer experiment}
Message transfer agar channel plates were prepared as follows; channel zones \emph{A}, \emph{B} and \emph{C} were removed from LB agar containing 30 $\mu$g/ul Kanamycin. Motility agar (Sigma-Aldrich) containing appropriate antibiotics were poured into channel zones A (25 $\mu$g/ml Streptomycin), B (100 $\mu$g/ml Ampicillin,12.5 $\mu$g/ml tetracycline, 0.5 \emph{mM} \emph{IPTG}) and \emph{C} (\emph{Streptomycin}, \emph{Ampicillin} and 0.5 \emph{mM} \emph{IPTG}) in turn. Strain \emph{HB101} was stab-inoculated to Zone \emph{A} and \emph{Novablue} containing \emph{pWITGLO} was stab-inoculated to Zone \emph{B} simultaneously. Plates were then incubated at 37\textdegree{}C for 3 days. 

\end{document}